\documentclass[a4paper,11pt]{article}
\pdfoutput=1 

\usepackage{jcappub} 

\usepackage[T1]{fontenc} 
\usepackage{mathrsfs}
\usepackage{tikz-feynman}
\usepackage{feynmf}
\usepackage{tikz}
\usepackage{slashed}

\newcommand{\ud}{\mathrm{d}}
\newcommand{\pd}{\partial}

\newcommand{\lie}{\mathscr L}

\newcommand{\order}[1]{\mathcal O\left(#1\right)}
\newcommand{\scri}{\mathscr I}

\newcommand{\ma}{\mathfrak a}
\newcommand{\mb}{\mathfrak b}
\newcommand{\mc}{\mathfrak c}
\newcommand{\md}{\mathfrak d}
\newcommand{\mn}{\mathfrak n}
\newcommand{\my}{\mathfrak y}

\newcommand{\hn}{\hat{n}}
\newcommand{\hnabla}{\hat{\nabla}}
\newcommand{\hf}{\hat{f}}
\newcommand{\hg}{\hat{g}}
\newcommand{\hL}{\hat{L}}
\newcommand{\hS}{\hat{S}}
\newcommand{\hR}{\hat{R}}
\newcommand{\hC}{\hat{C}}
\newcommand{\hK}{\hat{K}}
\newcommand{\hX}{\hat{X}}
\newcommand{\hth}{\hat{\vartheta}}
\newcommand{\hT}{\hat{T}}
\newcommand{\heps}{\hat{\epsilon}}

\newcommand{\bvec}[3]{(#1_{\hat{#2}})^{#3}}
\newcommand{\bform}[3]{(#1^{\hat{#2}})_{#3}}
\newcommand{\scon}[3]{\omega_{\hat{#1}}{}^{\hat{#2}}{}_{#3}}
\newcommand{\torsion}[2]{\mathcal T^{\hat{#1}}{}_{#2}}

\newcommand{\riem}[3]{\Omega_{#1\hat{#2}}{}^{\hat{#3}}}

\title{Conserved charges in Chern-Simons modified theory and memory effects}

\author[a]{Shaoqi Hou}
\author[b,c]{Tao Zhu}
\author[a,d]{Zong-Hong Zhu}

\affiliation[a]{School of Physics and Technology, Wuhan University,\\Wuhan, Hubei 430072, China}
\affiliation[b]{Institute for theoretical physics and cosmology, Zhejiang University of Technology,\\Hangzhou, Zhejiang 310032, China}
\affiliation[c]{United Center for Gravitational Wave Physics (UCGWP), Zhejiang University of Technology,\\Hangzhou, Zhejiang 310032, China}
\affiliation[d]{Department of Astronomy, Beijing Normal University,\\Beijing 100875,  China}

\emailAdd{hou.shaoqi@whu.edu.cn}
\emailAdd{zhut05@zjut.edu.cn}
\emailAdd{zhuzh@whu.edu.cn}

\abstract{
    In this work, conserved charges and fluxes at the future null infinity are determined in the asymptotically flat spacetime for Chern-Simons modified gravity.
    The flux-balance laws are used to constrain the memory effects.
    For tensor memories, the Penrose's conformal completion method is used to analyze the asymptotic structures and asymptotic symmetries, and then, conserved charges for the Bondi-Metzner-Sachs algebra are constructed with the Wald-Zoupas formalism.
    These charges take very similar forms to those in Brans-Dicke theory. 
    For the scalar memory, Chern-Simons modified gravity is rewritten in the first-order formalism, and the scalar field is replaced by a 2-form field dual to it. 
    With this dual formalism, the scalar memory is described by the vacuum transition induced by the large gauge transformation of the 2-form field.
}

\begin{document}
\maketitle
\flushbottom

\section{Introduction}

Gravitational memory effect was first discovered in the analysis of the gravitational wave from the close encounters of celestial bodies \cite{Zeldovich:1974gvh}.
It was represented as the permanent change in the metric perturbation long before and long after the encounter, and can be detected with laser interferometers \cite{Braginsky:1986ia}.
Earlier studies showed that this effect is given by the change in the quadrupole moment of the source of gravity \cite{1987Natur.327..123B}, but later, it was found out that the change in the stress-energy tensor of gravitational wave also results in memory effect \cite{Christodoulou1991,Wiseman:1991ss,Blanchet:1992br,Thorne:1992sdb}.
To differentiate, the former effect is called the linear or ordinary memory effect, and the later the nonlinear or null memory \cite{Bieri:2013ada}, which can actually be sourced by any null radiation.
Together, they are also called the displacement memory effect.
The memory effect has been calculated using post-Newtonian formalism for compact binary stars \cite{Favata:2008yd,Favata:2008ti,Favata:2010zu,Favata:2011qi}.

Recently, the memory effect is found to be closely related to the asymptotic symmetries of the isolated system \cite{Strominger2014bms,He:2014laa,Strominger:2014pwa}.
At the null infinity, the spacetime possesses the Bondi-Metzner-Sachs (BMS) symmetry, which generalizes the Poincar\'e symmetry by including supertranslations \cite{Bondi:1962px,Sachs:1962wk,Sachs:1962zza}.
Supertranslations can be roughly described as angle-dependent translations.
Because of them, the vacuum in the gravity sector is degenerate, and different vacuum states are related to each other via supertranslations. 
Transitions among vacua can occur due to the passage of the gravitational radiation.  
The change in vacuum is precisely the displacement memory effect \cite{Strominger:2014pwa}.
Its magnitude is constrained by the flux-balance law associated with the supertranslation \cite{Flanagan:2014kfa,Compere:2019gft}.
What is also interesting about this observation is the discovery of the infrared triangle, that is, the equivalence among the memory effect, soft graviton theorem \cite{Weinberg:1965s}, and supertranslation \cite{Strominger:2018inf}.

New devolvements also include the discovery of spin and center-of-mass (CM) memory effects \cite{Pasterski:2015tva,Nichols:2018qac}.
They both contribute to the so-called subleading displacement memory effect, which is present when the test particles initially move relative to each other. 
In contrast, the displacement memory, or more precisely, the leading displacement memory exists even if there is no initial relative velocity.
These new memories require the enlargement of the BMS symmetry, which contains super-rotation and super-boost transformations \cite{Barnich:2009se,Barnich:2010eb,Campiglia:2014yka,Campiglia:2015yka}. 
Spin memory is constrained by the flux-balance law associated with the super-rotation, while CM memory by the law for super-boost transformation \cite{Pasterski:2015tva,Flanagan:2015pxa,Nichols:2017rqr,Nichols:2018qac,Compere:2019gft}.
Spin memory, super-rotation symmetry and subleading soft graviton theorem constitute the three corners of a new triangular equivalence \cite{Cachazo:2014fwa,Kapec:2014opa,Campiglia:2014yka,Campiglia:2015yka,Pasterski:2015tva,Himwich:2019qmj}.
There are also more memories in general relativity \cite{Zhang:2017rno,Zhang:2017geq,Zhang:2018srn,Compere:2018ylh,Mao:2019sph,Chakraborty:2019yxn,Siddhant:2020gkn}. 
In electromagnetism and Yang-Mills theories, and even more theories, similar phenomena could also take place \cite{Bieri:2013hqa,Strominger:2018inf}.
But we will focus on displacement, spin and CM memories.

Since the gravitational wave has been detected \cite{Abbott:2016blz,TheLIGOScientific:2017qsa,LIGOScientific:2021qlt}, it is interesting to ask if the memory effect can be observed. 
Several works investigated the methods of and predicted the possibility of measuring the displacement memory with laster interferometers \cite{Lasky:2016knh,McNeill:2017uvq,Johnson:2018xly,Hubner2020mmn,Boersma:2020gxx,Liu:2021zys,Zhao:2021hmx}.
It can also be measured by pulsar timing arrays \cite{Seto:2009nv,Wang2015mm,Zhao:2021zlr} and the Gaia mission \cite{Madison:2020xhh}.
Spin memory effect is observable by LISA \cite{Pasterski:2015tva}, while CM memory is more difficult to be detected with the current and even the planed detectors \cite{Nichols:2018qac}.

Besides general relativity, there are alternative theories of gravity, proposed to at least partially resolve problems such as the breakdown of general relativity at the singularity, the nonrenormalizability, dark matter and dark energy, \textit{etc.}. 
In this work, we are interested in Chern-Simons modified gravity \cite{Jackiw:2003pm,Alexander:2009tp} and its memory effects.
In fact, previously, ref.~\cite{Hou:2021cbd} studied the asymptotically flat spacetimes in this theory with the Bondi-Sachs formalism, and the memory effect was analyzed. 
The asymptotically flat spacetime in this theory is similar to the one in general relativity, and thus possesses the same asymptotic symmetries \cite{Flanagan:2015pxa}.
If the Chern-Simons scalar field does not couple with ordinary matter fields, memory effects in the tensor sector are probably observable with laster interferometers, pulsar timing arrays and the Gaia mission.
These memories are similar to displacement, spin and CM memories in general relativity. 
They are constrained by properly integrating the evolutions equations of the Bondi mass and angular momentum aspects.
These constraints can also be written in terms of the flux-balance laws, which will be given in the current work. 
Moreover, the scalar memory exists, although it cannot be measured with the gravitational wave detectors.
To properly study it and its relation with symmetries, it is necessary to find a dual formalism in which the scalar field is dual to a 2-form field, which has the gauge symmetry. 
This allows one to interpret the relation between the scalar memory and the symmetry in the similar manner to the tensor memories.

For the purpose of determining the flux-balance laws, we will use Wald-Zoupas formalism \cite{Wald:1999wa}, suitably designed for any diffeomorphism invariant theory.
With this formalism, conserved charges and fluxes for the BMS symmetry can be calculated. 
Then, the flux-balance laws will be used to constrain the memory effects in the tensor sector. 
To facilitate the computation, the Penrose's conformal completion \cite{Penrose:1962ij,Penrose:1965am} is used, which has been applied to general relativity \cite{Wald:1999wa} and Brans-Dicke theory \cite{Brans:1961sx,Hou:2020wbo}.
To study the scalar memory, we will reformulate the original action according to ref.~\cite{Yoshida:2019dxu}, so that the Chern-Simons scalar is replaced by a 2-form field. 
The newly formulated action thus possesses the gauge transformation, which is nontrivial at the null infinity.
Like the displacement memory, the vacuum transition in the scalar sector, induced by the gauge transformation, causes the scalar memory.
The conserved charges and fluxes associated with the gauge transformation are obtained, and used to constrain the scalar memory. 
Since calculation with differential forms is very convenient, we will work in the first-order formalism. 
Conserved charges have also been calculated for certain spacetimes with Killing vectors in ref.~\cite{Tekin:2007rn}.

This work is organized as follows. 
In section~\ref{sec-cs}, Chern-Simons modified gravity will be briefly introduced, and the results in ref.~\cite{Hou:2021cbd} will be summarized.
Then, Penrose's conformal completion will be applied to analyze the asymptotic structure of the isolated system in Chern-Simons modified gravity, and the asymptotic symmetries in section~\ref{sec-cc}.
Section~\ref{sec-cops} will be devoted to the computation of conserved charges and fluxes to constrain memory effects in the tensor sector. 
The scalar memory effect will be discussed in section~\ref{sec-fir}.
Section~\ref{sec-con} is a brief summary.
The abstract index notation is used \cite{Wald:1984rg}.
Throughout this paper, $c=1$.
Most of the calculation was done with the help of \verb+xAct+ \cite{xact}.

\section{Chern-Simons modified gravity}
\label{sec-cs}

There are two basic methods to define an asymptotically flat spacetime in general relativity. 
The first one is the Bondi-Sachs formalism \cite{Bondi:1962px,Sachs:1962wk}, which has been generalized to some modified theories of gravity \cite{Hou:2020tnd,Tahura:2020vsa,Hou:2021cbd}. 
The second method is to use the Penrose conformal  completion \cite{Penrose:1962ij,Penrose:1965am}.
This section mainly reviews some properties of asymptotically flat spacetimes in Chern-Simons modified gravity, derived with Bondi-Sachs formalism \cite{Hou:2021cbd}.
The action is \cite{Alexander:2009tp}
\begin{equation}
    \label{eq-cs-act}
    S=\int\ud^4x\sqrt{-g}\left(\kappa R+\frac{\mathfrak a}{4}\vartheta R_{abcd}{}^*R^{bacd}-\frac{\mathfrak b}{2}\nabla_a\vartheta\nabla^a\vartheta\right),
\end{equation}
where $\kappa=1/16\pi G$, $\mathfrak a$ and $\mathfrak b$ are all constants.
$\nabla_a$ is the Levi-Civita connection.
$R_{abc}{}^d$ is the curvature tensor of $\nabla_a$, i.e., the Riemann tensor, ${}^*R^{bacd}=\epsilon^{cdef}R^{ba}{}_{ef}/2$ is its Hodge dual, and $R=g^{ab}R_{acb}{}^c$ the Ricci scalar.
The action is shift symmetric under the addition of a constant to $\vartheta$.
The presence of $\epsilon^{abcd}$ means that $\vartheta$ is a pseudo-scalar so that the action $S$ is invariant under the parity transformation.
$\ma$ and $\mb$ are free, so one may set $\ma\ne0$ and $\mb=0$. 
Then $\vartheta$ should be prescribed by hand, and the theory obtained is said to be \emph{nondynamical}.
When neither $\ma$ nor $\mb$ is zero, $\vartheta$ has its own dynamics. 
This framework is named \emph{dynamical}. 
In this paper, we work in the dynamical framework.
The equations of motion can be calculated via the variational principle \cite{Alexander:2009tp}
\begin{subequations}
    \label{eq-eoms}
\begin{gather}
    R_{ab}-\frac{1}{2}g_{ab}R+\frac{\mathfrak a}{\kappa}C_{ab}=\frac{1}{2\kappa}T_{ab}^{(\vartheta)},\label{eq-ein}\\
    \nabla_a\nabla^a\vartheta=-\frac{\mathfrak a}{4\mathfrak b}R_{abcd}{}^*R^{bacd}.\label{eq-eom-th}
\end{gather}
Here, $C_{ab}$ is called the C-tensor, given by 
\begin{equation}
    \label{eq-def-ct}
    C^{ab}=(\nabla_c\vartheta)\epsilon^{cde(a}\nabla_eR^{b)}{}_d+(\nabla_c\nabla_d\vartheta){}^*R^{c(ab)d},
\end{equation}
and $T_{ab}^{(\vartheta)}$ is the stress energy tensor of the Chern-Simons scalar field,
\begin{equation}
    \label{eq-def-ts}
    T^{(\vartheta)}_{ab}=\mathfrak b\left(\nabla_a\vartheta\nabla_b\vartheta-\frac{1}{2}g_{ab}\nabla_c\vartheta\nabla^c\vartheta\right).
\end{equation}
\end{subequations}

Usually, one uses Bondi-Sachs coordinates $(u,\,r,\,\theta,\,\phi)$ to describe an asymptotically flat spacetime, and the metric takes the following form,
\begin{equation}
    \label{eq-bc}
    \ud s^2=e^{2\beta}\frac{V}{r}\ud u^2-2e^{2\beta}\ud u\ud r+h_{AB}(\ud x^A-U^A\ud u)(\ud x^B-U^B\ud u),
\end{equation}
with $(x^2,\,x^3)=(\theta,\,\phi)$.
Here, $\beta,\,V,\,U^A$, and $h_{AB}$ ($A,B=2,3$) are six metric functions, satisfying
\begin{equation}
    \label{eq-exp-m}
    \beta=\order{r^{-1}},\quad V=-r+\order{r^0},\quad U^A=\order{r^{-2}},\quad
    \det(h_{AB})=r^4\sin^2\theta.
\end{equation}
Thus, $r$ is the luminosity radius.
One also requires that $\vartheta=\vartheta_0+\order{1/r}$, where $\vartheta_0$ is constant and set to zero due to the shift symmetry. 
With these conditions, one can solve eqs.~\eqref{eq-eoms} and find out that 
\begin{subequations}
    \label{eq-met-sols}
    \begin{gather}
        \vartheta=\frac{\vartheta_1}{r}+\frac{\vartheta_2}{r^2}+\order{\frac{1}{r^3}},\label{eq-sc-bc}\\
        V=-r+m+\order{\frac{1}{r}},\\
        \beta=\frac{\beta_2}{r^2}+\frac{\beta_3}{r^3}+\order{\frac{1}{r^4}},\quad \beta_2=-\frac{1}{32} c_{A}^{B} c^{A}_{B}-\frac{\mathfrak b}{16\kappa}\vartheta_1^2,\quad \beta_3=-\frac{\mb}{6\kappa}\vartheta_1\vartheta_2,\\
        U^A=-\frac{\mathscr D^B c_{AB}}{2r^2}+\frac{1}{r^3}\left[ -\frac{2}{3}N^A+\frac{1}{16}\mathscr D^A(c_{BC}c^{BC})+\frac{1}{2}c^{AB}\mathscr D^Cc_{BC} \right]+\order{\frac{1}{r^4}},\label{eq-user}\\
        h_{AB}=r^2\gamma_{AB}+rc_{AB}+\frac{1}{4}\gamma_{AB}c_C^Dc^C_D+\order{\frac{1}{r}},\label{eq-exp-h}
    \end{gather}
where all expansion coefficients are independent of $r$, $\gamma_{AB}$ is the round metric on a unit 2-sphere with $\gamma^{AB}$ its inverse, the indices are raised or lowered by $\gamma^{AB}$ or $\gamma_{AB}$, and $\mathscr D_A$ is the compatible covariant derivative of $\gamma_{AB}$.
In addition, $\gamma^{AB}c_{AB}=0$.
One also obtains the following evolution equations,
\begin{gather}
        \dot\vartheta_2=-\frac{1}{2}\mathscr D^2\vartheta_1,\label{eq-evo-th2}\\
        \dot m=\frac{1}{4}\mathscr D_A\mathscr D_BN^{AB}-\frac{1}{8}N_{AB}N^{AB}-\frac{\mb}{4\kappa}N^2,\label{eq-evo-m}\\
        \begin{split}
        \dot N_A=&\mathscr D_Am+\frac{1}{4}(\mathscr D_B\mathscr D_A\mathscr D_C c^{BC}-\mathscr D_B\mathscr D^B\mathscr D_C c_A^C)
        +\frac{1}{4}\mathscr D_C (c_{AB}N^{BC})+\frac{1}{2} c_{AB}\mathscr D_CN^{BC}\\
        &+\frac{\mb}{8\kappa}\left( \vartheta_1\mathscr D_AN-3N\mathscr D_A\vartheta_1 \right),\label{eq-evo-n}
        \end{split}
\end{gather}
\end{subequations}
where dot means to take the partial time derivative $\pd/\pd u$, $N=\dot\vartheta_1$, and $N_{AB}=\dot c_{AB}$ is the news tensor.
The last two equations are the evolution equations for the Bondi mass aspect $m$ and the angular momentum aspect $N_A$, respectively.
$N_{AB}$ and $N$ encode the information of the tensor and scalar gravitational waves, respectively. 
The vanishing of them implies the absence of gravitational waves near the null infinity.
In ref.~\cite{Hou:2021cbd}, eqs.~\eqref{eq-eoms} were solved at even higher orders in $1/r$, but for the purpose of the current work, it is sufficient to present these results.

Due to eq.~\eqref{eq-exp-m}, the spacetime possesses the BMS symmetry, generated by a vector field $\xi^a$, which parameterized by $\alpha(x^A)$ and $Y^A(x^B)$.
$\alpha$ generates a supertranslation. 
$Y^A$ is a global conformal Killing vector field on the unit 2-sphere, satisfying $\lie_Y\gamma_{AB}=\psi\gamma_{AB}$ with $\psi=\mathscr D_AY^A$.
So it generates Lorentz transformation.
The action of $\xi^a$ on $g_{ab}$ and $\vartheta$ is defined to be $\delta_\xi g_{ab}=\lie_\xi g_{ab}$ and $\delta_\xi\vartheta=\lie_\xi\vartheta$, respectively. 
Therefore, one has
\begin{subequations}
    \begin{gather}
    \delta_\xi\vartheta_1=fN+\frac{\psi}{2}\vartheta_1+\lie_Y\vartheta_1,    \label{eq-bms-phi1}\\
        \delta_\xi c_{AB}=fN_{AB}-2\mathscr D_A\mathscr D_Bf+\gamma_{AB}\mathscr D^2f+\lie_Y c_{AB}-\frac{\psi}{2}c_{AB},\label{eq-bms-c}
    \end{gather}
\end{subequations}
where the symbol $\lie_Y$ is to take the Lie derivative on the unit 2-sphere.
For more transformation rules, please refer to ref.~\cite{Hou:2021cbd}.
As mentioned in Introduction, one needs the enlarged BMS group to explain spin and CM memories. 
This means that $Y^A$ can be a Killing vector field on the unit 2-sphere with a finite number of singularities, so it belongs to the Virasoro algebra \cite{Blumenhagen:2009zz,Barnich:2009se,Barnich:2010eb}.
There is a second way to enlarge BMS group, that is, allowing $Y^A$ to generate all diffeomorphisms of the 2-sphere \cite{Campiglia:2014yka,Campiglia:2015yka}.
Then, $\gamma_{AB}$ changes under the action of $Y^A$, and, this leads to diverging symplectic current \cite{Flanagan:2019vbl}.
This divergence can be cured by renormalization using the ambiguity in the symplectic current \cite{Compere:2018ylh}. 
In this work, we will choose to work with the first kind of extension of the BMS algebra, as the computation of its charges and fluxes is easier. 

Since $\vartheta$ does not interact with the ordinary matter directly, and the second term (Chern-Simons coupling term) in action~\eqref{eq-cs-act} is of higher order in $1/r$,  the laser interferometer is incapable of detecting it, even if there exists the scalar gravitational wave. 
Of course, neither pulsar timing array nor the Gaia mission could detect it.
Therefore, with these gravitational wave detectors, one can observe memory effects of the tensor modes, i.e., displacement, spin and CM memories.
These tensor memory effects are very similar to those predicted in general relativity \cite{Strominger:2014pwa,Pasterski:2015tva,Flanagan:2015pxa,Strominger:2018inf,Nichols:2017rqr,Nichols:2018qac} and Brans-Dicke theory \cite{Hou:2020tnd,Tahura:2020vsa,Hou:2020wbo,Hou:2020xme,Tahura:2021hbk}.
In particular, they are constrained by the evolution equations \eqref{eq-evo-m} and \eqref{eq-evo-n}.
The constraints can also be written in terms of flux-balance laws, which will be derived below.

\section{Asymptotically flat spacetime in conformal completion formalism}
\label{sec-cc}

In the previous section, the Bondi-Sachs formalism for Chern-Simons modified gravity has been briefly reviewed, so in the current section, the Penrose conformal completion will be developed. 
This method will also provide the asymptotic structure defined at null infinity \cite{Ashtekar:1981bq,Ashtekar:2014zsa}, and BMS symmetries.
These will be useful for computing the conserved charges and fluxes in section~\ref{sec-cops}.

As in general relativity and Brans-Dicke theory \cite{Wald:1984rg,Hou:2020wbo}, the asymptotically flat spacetime at $\scri$ in Chern-Simons modified gravity is a spacetime $(M,g_{ab})$ satisfying the following conditions,
\begin{enumerate}
    \item There exists an unphysical spacetime $(\hat M,\hat g_{ab})$ and a conformal transformation $\mathcal C:M\rightarrow\mathcal C[M]\subset \hat M$ such that 
    \begin{equation}
        \label{eq-def-cf}
        \hat g_{ab}=\Omega^2\mathcal C^*g_{ab},
    \end{equation}
    for some conformal factor $\Omega$ with $\mathcal C^*$ the pullback map;

    \item $\scri$ is the boundary of $M$ in $\hat M$, and on it, $\Omega=0$ and $\nabla_a\Omega\ne0$;
    
    \item the topology of $\scri$ is $\mathbb S^2\times\mathbb R$;
    
    \item equations~\eqref{eq-eoms} are satisfied near $\scri$.
\end{enumerate}
Although $\vartheta$ does not define the spacetime geometry directly, we also want to define its conformal transformation, i.e.,
\begin{equation}
    \label{eq-def-cft}
    \hat\vartheta=\Omega^{-1}\vartheta,
\end{equation}
where $\hat\vartheta$ is the scalar field in the unphysical spacetime.

\subsection{Asymptotic structure at null infinity}
\label{sec-asystr}

Now, in the unphysical spacetime, the equations of motion become 
\begin{subequations}
    \label{eq-eoms-up}
    \begin{gather}
        \Omega\hat S_{ab}+2\hat\nabla_a\hat n_b-\hat f\hat g_{ab}=\Omega^{-1}\hat L_{ab},\label{eq-eoms-s-c}\\
        \Omega\hat\nabla_a\hat\nabla^a\hat\vartheta+\hat\vartheta\hat\nabla_a\hat n^a-2\hat f\hat\vartheta=-\frac{\ma}{4\mb}\Omega^2\hat R_{abcd}{}^*\hat R^{bacd},\label{eq-eoms-th-c}
    \end{gather}
where $\hS_{ab}=\hR_{ab}-\hg_{ab}\hR/6$ is the Schouten tensor for $\hg_{ab}$, $\hn_a=\hnabla_a\Omega$, $\hf=\hn_a\hn^a/\Omega$.
The effective stress-energy tensor $\hL_{ab}$ in the unphysical spacetime is
\begin{equation}
    \label{eq-def-hL}
    \hL_{ab}=\frac{\Omega^2}{2\kappa} \left( \hT_{ab}^{(\vartheta)}-\frac{1}{6}\hg_{ab}\hT^{(\vartheta)} \right)+\frac{\ma\Omega^3}{\kappa}\hat C_{ab},
\end{equation}
with 
\begin{gather}
    \hT^{(\vartheta)}_{ab}=\hth^2\hn_a\hn_b+2\Omega\hn_{(a}\hnabla_{b)}\hth+\Omega^2\hnabla_a\hth\hnabla_b\hth,\quad \hT^{(\vartheta)}=\hg^{ab}\hT^{(\vartheta)}_{ab},\\
    \begin{split}
        \hat C_{ab}=&6\hth {}^*\hR_{acbd}\hn^c\hn^d+\Omega \left[ 7 {}^*\hR_{acbd}\hn^{(c}\hnabla^{d)}\hth + {}^*\hR_{acbd} \hnabla^{(d}(\hn^{c)}\hth)+\hn^c\hat\epsilon_{cde(a}\hS_{b)}^d\hnabla^e\hth\right.\\
        &\left.+\hth\hn^c\hat\epsilon_{cde(a}\hnabla^d\hS_{b)}^e\right]+\Omega^2 \left(  {}^*\hR_{acbd}\hnabla^c\hnabla^d\hth+\hat\epsilon_{cde(a}\hnabla^d\hS_{b)}^e\hnabla^c\hth \right).
    \end{split}
\end{gather}
\end{subequations}
In these expressions, all hatted quantities are defined in the unphysical spacetime, they are smooth at $\scri$, and their indices are raised and lowered by $\hg_{ab}$ and $\hg^{ab}$, respectively.
At this point, one should compare these relations to eqs.~(13), (14) and (15) in ref.~\cite{Hou:2020wbo}.
Then, one immediately finds out that eqs.~\eqref{eq-eoms-up} become eqs.~(13), (14) and (15), modulo certain coupling constants, if one ignores the terms containing Riemann tensor and Schouten tensor on the right-hand sides.
In fact, these extra terms  are of higher orders in $\Omega$, as one approaches $\scri$ with $\Omega\rightarrow0$.
Therefore, many conclusions below resemble those in Brans-Dicke theory \cite{Hou:2020wbo}.
For example, the smoothness of these equations near $\scri$ implies that $\hn_a$ is a null covector field on $\scri$. 
Since $\hn_a=\hnabla_a\Omega$, and $\Omega$ is constant on $\scri$, $\hn_a$ is a null normal to $\scri$, so $\scri$ is a null hypersurface as it should be.

The conformal factor $\Omega$ can be freely chosen. 
Let $\varpi$ be a positive smooth function near $\scri$, then $\Omega'=\varpi\Omega$ is also a conformal factor.
Under such kind of gauge transformation, one can check that 
\begin{subequations}
    \begin{gather}
        \hg'_{ab}=\varpi^2\hg_{ab},\quad \hth'=\varpi^{-1}\hth,\quad \hn'_a=\varpi\hn_a+\Omega\hnabla_a\varpi,\\
        \hf'=\varpi^{-1}\hf+2\varpi^{-2}\hn^a\hnabla_a\varpi+\varpi^{-3}\Omega(\hnabla^a\varpi)\hnabla_a\varpi.
    \end{gather}
\end{subequations}
One usually chooses Bondi gauge where $\hf=0$ on $\scri$.
This also implies $\hnabla_a\hn_b=0$ on $\scri$ by eqs.~\eqref{eq-eoms-up}. 
In the following, the symbol $\ddot =$ indicates to evaluate equations on $\scri$, so in Bondi gauge, $\hf\ddot =0$ and $\hnabla_a\hn_b\ddot=0$.
These imply that the integral curves of $\hn^a$ on $\scri$ are generators of $\scri$, and form null geodesic congruence with vanishing convergence, shear and rotation \cite{Wald:1984rg}.
Equation~\eqref{eq-eoms-s-c} is evaluated to 
\begin{equation}
    \label{eq-nk}
    \lie_{\hn}\hg_{ab}\ddot=0,
\end{equation}
so $\hn^a$ is also a Killing vector field on $\scri$.
The residual gauge transformation is permitted with $\hn^a\hnabla_a\varpi=0$.
From now on, Bondi gauge will be imposed to simplify the discussion.

Up to now, one has already identified the zeroth-order structure $(\hg_{ab},\hn^a,\hth)$ on $\scri$ \cite{Ashtekar:2014zsa,Hou:2020wbo}.
Since these tensors are defined on $\hat M$, one may want to obtain the corresponding quantities that are intrinsic to $\scri$.
So let $\gamma_{ab}$ be the induced metric of $\hg_{ab}$ on $\scri$. 
Then, the zeroth-order structure is $(\gamma_{ab},\hn^a,\hth)$, where $\hn^a$ is naturally an intrinsic vector field on $\scri$ since it is tangent to $\scri$.
It is worth to note that $\gamma_{ab}\hn^b=0$ as $\hg_{ab}\hn^b=0$.

A foliation of $\scri$ can be determined with the zero-order structure. 
Assume the generators of $\scri$ are affinely parameterized by $u$ so that $\hn^a=(\ud/\ud u)^a$.
Let $\mathscr C_0$ be a cross section of $\scri$, that is, a topological 2-sphere on $\scri$ and each generator of $\scri$ punching through it only once.
One can choose a suitable parameterization of the generators of $\scri$ such that on $\mathscr C_0$, $u=u_0$.
By Lie-dragging $\mathscr C_0$ along the generators of $\scri$, one obtains a foliation of $\scri$.

The spacetime covariant derivative $\hnabla_a$ induces a covariant derivative $\mathscr D_a$ that is intrinsic to $\scri$, and is the first-order structure on $\scri$.
It is compatible with $\gamma_{ab}$,  and annihilates $\hn^a$,
\begin{equation}
    \label{eq-copro}
    \mathscr D_c\gamma_{ab}=0,\quad\mathscr D_a\hn^b=0.
\end{equation}
With $\mathscr D_a$, one defines its Riemann tensor $\mathscr R_{abc}{}^d$ in the usual manner, i.e., for any covector field $\nu_a$ on $\scri$, $(\mathscr D_a\mathscr D_b-\mathscr D_b\mathscr D_a)\nu_c=\mathscr R_{abc}{}^d\nu_d$.
Let $\mathscr R_{abcd}=\gamma_{de}\mathscr R_{abc}{}^e$, then, $\mathscr R_{ab}=\gamma^{cd}\mathscr R_{acbd}$ and $\mathscr R=\gamma^{ab}\mathscr R_{ab}$ are the Ricci tensor and scalar, respectively.
Here, $\gamma^{ab}$ is the inverse of $\gamma_{ab}$ such that $\gamma_{ac}\gamma_{bd}\gamma^{cd}=\gamma_{ab}$.
As shown in ref.~\cite{Ashtekar:1981hw}, $\mathscr R_{abc}{}^d=\gamma_{c[a}\mathcal S_{b]}{}^d+\mathcal S_{c[a}\delta_{b]}^d$ with $\mathcal S_{ab}=\gamma_{bc}\mathcal S_a{}^c$, for some tensor field $\mathcal S_a{}^b$ satisfying $\mathcal S_b{}^a\hn^b=(\mathcal S_b{}^b-\mathscr R)\hn^a$.
$\mathcal S_a{}^b$ is just the restriction of $\hS_a{}^b$ to $\scri$.

Now, one can introduce the second-order structure on $\scri$. 
Since $\scri$ has the topology of a cylinder, a unique symmetric tensor field $\rho_{ab}$ can be determined satisfying the following conditions \cite{Geroch1977},
\begin{equation}
    \label{eq-def-rho}
    \rho_{ab}\hn^b=0,\quad \gamma^{ab}\rho_{ab}=\mathscr R,\quad \mathscr D_{[a}\rho_{b]c}=0.
\end{equation}
In fact, $\rho_{ab}$ can be $\mathscr R\gamma_{ab}/2$ in a gauge with $\mathscr R$ constant.
With it, the news tensor $N_{ab}$ can be defined,
\begin{equation}
    \label{eq-def-news}
    N_{ab}=\rho_{ab}-\mathcal S_{ab}.
\end{equation}
Together with $N=\lie_{\hn} \hth$, $N_{ab}$ forms the second-order structure on $\scri$.

The third-order structure is constructed out of the Weyl tensor $\hC_{abcd}$ of $\hnabla_a$.
By Theorem 11 in ref.~\cite{Geroch1977}, one can still show that $\hC_{abcd}=0$ on $\scri$, although the energy-momentum tensor $\hL_{ab}$ contains more terms than the one in ref.~\cite{Geroch1977}. 
So, one can define 
\begin{equation}
    \label{eq-def-Ks}
    \hK^{ab}=-4\Omega^{-1}\hC^{acbd}\hn_c\hn_d,\quad {}^*\hK^{ab}=-4\Omega^{-1}{}^*\hC^{acbd}\hn_c\hn_d.
\end{equation}
These are naturally defined on $\scri$, and the third-order structure.
After some complicated computation, one finds out that
\begin{subequations}
    \begin{gather}
        \mathscr D_{[a}\mathcal S_{b]}{}^c=\frac{1}{4}\heps_{abd}{}^*\hK^{dc},\label{eq-shk-div}\\
        \mathscr D_b\hK^{ab}=\frac{2\mb}{3\kappa}\hn^a(\hth\lie_{\hn} N-2 N^2),\label{eq-hk-div}\\
         \mathscr D_b{}^*\hK^{ab}=0,
    \end{gather}
\end{subequations}
in Bondi gauge, where $\heps_{abc}$ is the induced volume element on $\scri$, i.e., $\heps_{abcd}=4\heps_{[abc}\hn_{d]}$.

Under the gauge transformation $\Omega\rightarrow\varpi\Omega$, these structures transform in the following way,
\begin{subequations}
    \begin{gather}
        \gamma'_{ab}=\varpi^2\gamma_{ab},\quad \hn'^a=\varpi^{-1}\hn^a, \quad \hth'=\varpi^{-1}\hth,\\
        \mathscr D'_a\nu_b=\mathscr D_a\nu-2\varpi^{-1}\nu_{(a}\mathscr D_{b)}\varpi+\varpi^{-1}\gamma_{ab}\varpi^c\nu_c,\\
        N'_{ab}=N_{ab},\quad N'=\varpi^{-2}N,\\
        \hK'^{ab}=\varpi^{-5}\hK^{ab},\quad {}^*\hK'^{ab}=\varpi^{-5}{}^*\hK^{ab},
    \end{gather}
\end{subequations}
where $\varpi^a$ is the restriction of $\hnabla^a\varpi$ to $\scri$.
From these transformation laws, one finds out that $N_{ab}$ is special, as it is invariant.
$\mathscr D_a$ is even more special in the following sense. 
Consider $\varpi=1+\Omega \varpi'$ for some smooth function $\varpi'$ near $\scri$, then one has
\begin{equation}
    \label{eq-dt1}
    \mathscr D'_a\nu_b=\mathscr D_a\nu_b+\varpi'\gamma_{ab}\hn^c\nu_c,
\end{equation}
while the remaining structures stay the same.
This relation suggests to introduce a new tensor operator $\{\mathscr D_a\}$, the equivalence class of $\mathscr D_a$, which is the set of $\mathscr D_a$'s different from each other according to eq.~\eqref{eq-dt1} \cite{Ashtekar:1981hw}.
The radiative degrees of freedom in the tensor sector are related to $\{\mathscr D_a\}$. 
To see this, consider two unphysical metrics $\hg_{ab}$ and $\hg'_{ab}$\footnote{These two metrics are not necessarily conformally related.}, and their respective covariant derivatives $\hnabla_a$ and $\hnabla'_a$.
These covariant derivatives naturally induce $\{\mathscr D_a\}$ and $\{\mathscr D'_a\}$, respectively.
They differ from each other by a symmetric tensor $\sigma_{ab}$ with $\sigma_{ab}\hn^b=0$ and $\gamma^{ab}\sigma_{ab}=0$. 
In fact, it can be shown that  \cite{Ashtekar:1981hw}
\begin{equation}
(\mathscr D'_a-\mathscr D_a)\nu_b=\Sigma_{ab}\hn^c\nu_c,\quad \sigma_{ab}=\Sigma_{ab}-\frac{1}{2}\gamma_{ab}\gamma^{cd}\Sigma_{cd},   \label{eq-def-sig-0}
\end{equation}
where $\hn^c\Sigma_{ab}$ is the connection, and $\mathscr D'_a$ and $\mathscr D_a$ are the representatives of $\{\mathscr D'_a\}$ and $\{\mathscr D_a\}$, respectively.
More specifically, let $\ell_a$ be a covector field on $\scri$ such that $\hn^a\ell_a=1$, then $\sigma_{ab}$ is just the traceless part of $\Sigma_{ab}=(\mathscr D'_a-\mathscr D_a)\ell_b$.
Assume that there exists a particular covariant derivative $\mathring{\mathscr D}_a$ with $\mathring{\mathscr D}_a\ell_b=0$ \cite{Ashtekar:1981hw}, then we specifically define 
\begin{equation}
    \label{eq-def-sig}
\Sigma_{ab}=(\mathscr D_a-\mathring{\mathscr D}_a)\ell_b=\mathscr D_a\ell_b,
\end{equation}
from now on.
With these, it can be shown that $N_{ab}=2\lie_{\hn}\sigma_{ab}$ \cite{Hou:2020wbo}.
$\sigma_{ab}$ is the shear of a null congruence with tangent vector fields $\ell^a=\hg^{ab}\ell_b$ on $\scri$.
Usually, one  sets $\ell_a=\mathscr D_au$.

Let us now identify the vacuum in the tensor sector. 
Among all the asymptotic structures, the zeroth-structure $(\gamma_{ab},\hn^a)$ is universal, that is, any asymptotically flat spacetime possesses it \cite{Geroch1977,Flanagan:2019vbl}.
The first-order structure $\{\mathscr D_a\}$ can be different for different asymptotically flat spacetime, so some particular $\{\mathscr D_a\}$ defines the vacuum.
According to ref.~\cite{Ashtekar:1981hw}, if the curvature of $\{\mathscr D_a\}$ is trivial, the spacetime is said to be in the vacuum.
This implies that $N_{ab}$ and ${}^*\hK^{ab}$ are zero, which means that there is no tensor gravitational wave, and the spacetime is also said to be nonradiative.
Of course, the condition of being in vacuum can be simply ${}^*\hK^{ab}=0$, as this leads to the vanishing of $N_{ab}$ via eqs.~\eqref{eq-def-news} and \eqref{eq-shk-div}.
In terms of $\sigma_{ab}$, a vacuum state is characterized by
\begin{equation}
    \label{eq-def-vac-t}
    \sigma_{ab}=\frac{1}{2}\mathscr D_a\mathscr D_b \Phi-\frac{1}{4}\gamma_{ab}\mathscr D^2\Phi,
\end{equation}
for some function $\Phi$ on $\scri$ with $\lie_{\hn}\Phi=0$ \cite{Ashtekar:1981hw}.
The vacuum state in the scalar tensor is simply given by $N=0$.

In terms of the quantities defined in section~\ref{sec-cs}, one can check that $\Omega=1/r$, $\hth=\vartheta_1$, and $\hn^a=(\pd_u)^a$. 
In addition, $\gamma_{ab}$ and $\sigma_{ab}$ have nonvanishing components in the angular directions, i.e., $(\pd_A)^a(\pd_B)^b\gamma_{ab}=\gamma_{AB}$, and $(\pd_A)^a(\pd_B)^b\sigma_{ab}=c_{AB}/2$. 
So, $N_{AB}=(\pd_A)^a(\pd_B)^bN_{ab}=\pd_uc_{AB}$, which explains why eq.~\eqref{eq-def-news} has a sign difference from the usual definition \cite{Ashtekar:1981hw}.
Interestingly, the vacuum is given by 
\begin{equation}
    \label{eq-def-vac}
c_{AB}=\mathscr D_A\mathscr D_B\Phi-\frac{1}{2}\gamma_{AB}\mathscr D^2\Phi,
\end{equation}
which is consistent with the definition in ref.~\cite{Hou:2021cbd}.

\subsection{Infinitesimal BMS symmetries}

In a spacetime with Killing vector fields $K^a$, the metric field $g_{ab}$ is preserved by the diffeomorphisms generated by $K^a$, so $\lie_Kg_{ab}=0$.
Such a relation may not exist in a general spacetime. 
For the asymptotically flat spacetime considered here, there may not exist Killing vector fields, either, but it is possible that $\lie_\xi g_{ab}$ approaches zero at an appropriate rate for some vector field $\xi^a$.
If such $\xi^a$ exists, one states that it generates the asymptotic symmetry. 
More precisely, one requires the following relation hold \cite{Geroch:1981ut},
\begin{equation}
    \label{eq-def-bms}
    \Omega^2\delta_\xi g_{ab}\equiv \Omega^2\lie_{\xi}g_{ab}=\lie_\xi\hg_{ab}-\psi\hg_{ab}=2\Omega\hX_{ab},
\end{equation}
for some smooth function $\psi=2\xi^a\hn_a/\Omega$, and some smooth tensor field $\hX_{ab}$ near $\scri$.
The smoothness of $\psi$ implies $\xi^a\hn_a\ddot=0$, so $\xi^a$ is tangent to $\scri$.
Usually, one reexpresses eq.~\eqref{eq-def-bms} in the following way,
\begin{equation}
    \label{eq-def-bms-1}
    \lie_\xi\hg_{ab}=\psi\hg_{ab}+2\Omega\hX_{ab},
\end{equation}
which means that $\xi^a$ is a conformal Killing vector field for $\hg_{ab}$ on $\scri$ (where $\Omega=0$).
Since $\hn^a=\hg^{ab}\hnabla_b\Omega$, one knows that 
\begin{equation}
    \label{eq-lien}
    \lie_\xi\hn^a=-\frac{\psi}{2}\hn^a+\Omega\hnabla^a\frac{\psi}{2}-2\Omega^2\hX^a,
\end{equation}
where $\hX^a=\hg^{ab}\hX_{bc}\hn^c/\Omega$.
Further contracting both sides of $\hn_a$ leads to 
\begin{equation}
    \label{eq-liepsi}
    \lie_{\hn}\psi=\frac{1}{4}\left(\lie_\xi\hf-\frac{1}{2}\psi\hf\right),
\end{equation}
without imposing Bondi gauge condition.
So in Bondi gauge, $\lie_{\hn}\psi=0$.
Finally, the action of $\xi^a$ on $\hth$ is 
\begin{equation}
    \label{eq-lieth}
    \delta_\xi\hth=\lie_\xi\hth+\frac{\psi}{2}\hth.
\end{equation}
This is consistent with \eqref{eq-bms-phi1}.

Now, restricting eqs.~\eqref{eq-def-bms-1}, and \eqref{eq-lien} to $\scri$, one obtains how an infinitesimal BMS transformation acts on the zeroth-order asymptotic structure,
\begin{equation}
    \label{eq-lie-scri}
    \delta_\xi\gamma_{ab}=\lie_\xi\gamma_{ab}-\psi\gamma_{ab}=0,\quad\delta_\xi\hn^a= \lie_\xi\hn^a+\frac{\psi}{2}\hn^a=0,
\end{equation}
although they look trivial.
The action of $\xi^a$ on $\hth$ is just eq.~\eqref{eq-lieth} on $\scri$.
Next, consider the action of $\xi^a$ on the first-order asymptotic structure $\mathscr D_a$, given by 
\begin{equation}
    \label{eq-lie-cod}
    (\lie_\xi\mathscr D_a-\mathscr D_a\lie_\xi)\nu_b=(\xi^d\mathscr R_{dab}{}^c-\mathscr D_a\mathscr D_b\xi^c)\nu_c,
\end{equation}
using the definitions of the Lie derivative and the Riemann tensor.
Now, let $\nu_a=\ell_a$, then one obtains the variation of $\Sigma_{ab}$
\begin{equation}
    \label{eq-liesig}
    \delta_\xi\Sigma_{ab}=-\mathscr D_a\mathscr D_b(\xi^c\ell_c)+2\sigma_{c(a}\mathscr D_{b)}\xi^c+\xi^c\mathscr D_c\sigma_{ab}+\frac{\psi}{4}\ell\gamma_{ab}+\frac{1}{2}\gamma_{ab}\xi^c\mathscr D_c\ell,
\end{equation}
where $\ell=\gamma^{ab}\mathscr D_a\ell_b$.
The traceless part of this equation gives the transformation of $\sigma_{ab}$.
In particular, the supertranslation changes $\sigma_{ab}$ by 
\begin{equation}
    \label{eq-sup-tf-s}
    \delta_\alpha\sigma_{ab}=\frac{\alpha}{2}N_{ab}-\mathscr D_a\mathscr D_b\alpha+\frac{1}{2}\gamma_{ab}\mathscr D^2\alpha.
\end{equation} 
If one applies this transformation to a vacuum state described by eq.~\eqref{eq-def-vac-t}, one will get a new state $\sigma'_{ab}=\sigma_{ab}+\delta_\alpha\sigma_{ab}$, which is also vacuum with $\Phi'=\Phi-2\alpha$.
This observation explains the displacement memory to be discussed later.

Another useful expression for the left-hand side of eq.~\eqref{eq-lie-cod} can be obtained by considering the change in $\hnabla_a$ in the bulk manifold $\hat M$.
The infinitesimal diffeomorphism $\xi^a$ changes $\hg_{ab}$ according to \eqref{eq-def-bms}, so the corresponding Levi-Civita connection changes by $\delta_\xi\hat\Gamma^c{}_{ab}$.
One thus has 
\begin{equation*}
    \label{eq-lie-cod-s}
    (\lie_\xi\hnabla_a-\hnabla_a\lie_\xi)\nu_b=-\nu_c\delta_\xi\hat\Gamma^c{}_{ab}=\nu_c\left[\hn^c\hX_{ab}-2\hn_{(a}\hX_{b)}{}^c-\delta^c_{(a}\hnabla_{b)}\psi+\hg_{ab}\hnabla^c\frac{\psi}{2}+\order{\Omega}\right].
\end{equation*}
Restricting this equation to $\scri$ and setting $\nu_a=\ell_a$, one obtains the following useful relation,
\begin{equation}
    \label{eq-sec-xab}
    \hX_{ab}=(\lie_\xi\mathscr D_a-\mathscr D_a\lie_\xi)\ell_b+\ell_{(a}\mathscr D_{b)}\psi-\gamma_{ab}\ell_c\hnabla^c\frac{\psi}{2},
\end{equation}
where $\hnabla^c$ is not replaced by $\mathscr D_a$, since the last term is not useful for calculating fluxes.

Now, one can write down the form of $\xi^a$ more explicitly.
Since $\hn^a$ is tangent to $\scri$, it can be a BMS generator. 
In fact, a generic vector field $\xi^a\ddot=\alpha \hn^a$ is also a BMS generator, if $\lie_{\hn}\alpha\ddot=0$.
Such kind of vector field generates the so-called supertranslation transformation.
One can show that $\psi=0$, so $\alpha\hn^a$ is a Killing vector field for $\gamma_{ab}$.
A generic BMS generator $\xi^a$ can be expressed in the following manner,
\begin{equation}
    \label{eq-xi-dec}
    \xi^a\ddot= \left( \alpha+\frac{u}{2}\psi \right)\hn^a+Y^a,
\end{equation}
using the foliation of $\scri$.
Here, $Y^a$ is tangent to the  cross section $\mathscr C$ at $u=0$, and the conformal Killing vector field $\lie_Y\gamma_{ab}=\psi\gamma_{ab}$ with $\psi=\mathscr D_aY^a$.
So $Y^a$ generates Lorentz transformation on $\mathscr C$.
$Y^a$ can be extended to all $\scri$ by Lie-dragging $\lie_{\hn}Y^a=0$ \cite{Hou:2020wbo}.
The set of $\xi^a$'s is just the BMS algebra, with the following Lie bracket 
\begin{equation}
    \label{eq-def-lib}
    [\xi_1,\xi_2]^a\ddot=\left[\lie_{Y_1}\alpha_2-\lie_{Y_2}\alpha_1+\frac{u}{2}(\lie_{Y_1}\psi_2-\lie_{Y_2}\psi_1)  \right]\hn^a+\lie_{Y_1}Y_2^a,
\end{equation}
where $\mathscr D_a(\lie_{Y_1}Y^a)=\lie_{Y_1}\psi_2-\lie_{Y_2}\psi_1$.
From this, one can find out that supertranslation generators  constitute the abelian ideal of the BMS algebra, and thus, the BMS algebra is the semi-direct sum of the supertranslation algebra and the Lorentz algebra.

In section~\ref{sec-cs}, the BMS transformation is also reviewed. 
There, the supertranslation generator is labeled by $\alpha$, as done in this section. 
$Y^A$ represents the angular component of $Y^a$, i.e., $Y^A=Y^a(\ud x^A)_a$.
Rewriting eq.~\eqref{eq-liesig} in Bondi-Sachs coordinates, one recovers eq.~\eqref{eq-bms-c},
and eq.~\eqref{eq-lieth} is exactly eq.~\eqref{eq-bms-phi1} as $\hth=\vartheta_1$.

\section{Conserved charges for BMS algebra and memory effects}
\label{sec-cops}

The conserved charges and fluxes will be calculated with the Wald-Zoupas formalism \cite{Wald:1999wa} for BMS algebra in this section.
Then, the flux-balance laws are used to constrain the memory effects in the tensor sector.

\subsection{Covariant phase space}

Following ref.~\cite{Wald:1999wa}, the presymplectic potential current for Chern-Simons modified gravity is,
\begin{equation}
    \label{eq-psp}
    \theta_{abc}[\delta]=\epsilon_{dabc}(\kappa v^d_\text{GR}+v^d_\vartheta+u^d[\delta]),
\end{equation}
with $\delta$ representing field variations $(\delta g_{ab},\delta\vartheta)$ collectively.
Here, $v_\text{GR}^a$ comes from $\delta R$, $v^a_\vartheta$ from the variation of the third term in action~\eqref{eq-cs-act}, and $u^a$ from the variation of the coupling term, which are 
\begin{subequations}
    \begin{gather}
        v^a_\text{GR}=g^{ab}g^{cd}(\nabla_c\delta g_{bd}-\nabla_b\delta g_{cd}),\\
        v_\vartheta^a=-\mb\delta\vartheta\nabla^a\vartheta,\\
        u^a[\delta]=\ma[\vartheta{}^*R^{cbad}\nabla_c\delta g_{bd}-\nabla_c(\vartheta{}^*R^{abcd})\delta g_{bd}],
    \end{gather}
\end{subequations}
respectively.
The presymplectic current 3-form can thus be calculated,
\begin{equation}
    \label{eq-psc}
    \omega_{abc}[\delta,\delta']=\delta\theta_{abc}[\delta']-\delta'\theta_{abc}[\delta]=\epsilon_{dabc}(\kappa w^d_\text{GR}+w^d_\vartheta+z^d),
\end{equation}
where $\delta'$ stands for $(\delta'g_{ab},\delta'\vartheta)$, and \cite{Hou:2020wbo}
\begin{subequations}
    \begin{gather}
        w^a_\text{GR}=(g^{a[e}g^{d]c}g^{bf}+g^{ae}g^{b[f}g^{c]d}+g^{a[d}g^{b]c}g^{ef})(\delta'g_{bc}\nabla_d\delta g_{ef}-\delta g_{bc}\nabla_d\delta'g_{ef}),\\
        w^a_\vartheta=-\mb \left[ \delta'\vartheta\nabla^a\delta\vartheta+\delta'\vartheta\delta g^{ab}\nabla_b\vartheta+\frac{g^{bc}\delta g_{bc}}{2}\delta'\vartheta\nabla^a\vartheta-\langle\delta\leftrightarrow\delta'\rangle \right],\\
        \begin{split}
            z^a=&\ma \bigg[-{}^*R^{acbd}\delta'g_{cd}\nabla_b\delta\vartheta-{}^*R^{bdac}\delta\vartheta\nabla_d\delta'g_{bc}+\nabla_b\vartheta \left( \frac{1}{2}{}^*R^{acbd}g^{ef}\delta g_{ef}\right.\\
            &\left.-\frac{1}{2}\epsilon^{bdfh}R^{ace}{}_h\delta g_{ef}+2g^{f[c}{}^*R^{a]ebd}\delta g_{ef} -g^{a[p}g^{q]c}\epsilon^{bdef}\nabla_p\nabla_f\delta g_{eq} \right)\delta'g_{cd}\\
            &+\vartheta \left( \epsilon^{ace}{}_fg^{bh}\nabla_{[d}\nabla^f\delta g_{b]e}\nabla^d\delta'g_{ch} -\frac{1}{2}{}^*R^{cdae}\delta g_{bc}\nabla^b\delta'g_{de}\right.\\
            &\left.+\frac{1}{2}\epsilon^{ace}{}_hR^{bhdf}\delta g_{bc}\nabla_f\delta'g_{de}+g^{bf}{}^*R^{cead}\delta g_{bc}\nabla_{[e}\delta'g_{d]f}\right.\\
            &\left.+\frac{1}{2}g^{bf}{}^*R^{cead}\delta g_{bf}\nabla_e\delta'g_{cd}\right)\bigg]
            +\frac{1}{2}u^a[\delta']g^{bc}\delta g_{bc}-\langle\delta\leftrightarrow\delta'\rangle,
        \end{split}
    \end{gather}
\end{subequations}
with $\langle\delta\leftrightarrow\delta'\rangle$ meaning the terms obtained by exchanging $\delta$ and $\delta'$ of the previous terms.
As one can guess, $\kappa w^a_\text{GR}+w^a_\vartheta$ takes the similar form to that in Brans-Dicke theory \cite{Hou:2020wbo}, and $z^a$ comes from the coupling term.

Now, one can calculate the presymplectic form $\Xi_\Sigma$, which is 
\begin{equation}
    \label{eq-def-xi}
    \Xi_\Sigma[\delta;\delta']=\int_\Sigma\omega_{abc}[\delta,\delta'],
\end{equation}
for an arbitrary, closed, embedded 3-dimensional hypersurface $\Sigma$ without boundary.
If the variation $\delta'$ is induced by a vector field $\xi^a$, that is, $\delta'_\xi g_{ab}=\lie_\xi g_{ab}$ and $\delta'_\xi\vartheta=\lie_\xi\vartheta$, then eq.~\eqref{eq-def-xi} gives the variation of a Hamiltonian, or a charge, $Q_\xi$, conjugate to $\xi^a$ \cite{Wald:1999wa},
\begin{equation}
    \label{eq-def-ham}
    \slashed\delta Q_\xi[\Sigma]=\int_\Sigma\omega_{abc}[\delta,\lie_\xi],
\end{equation}
provided that the equations of motion and the linearized equations of motion are all satisfied.
For the diffeomorphism invariant theories of gravity, this hypersurface integral can be rewritten as the one over a 2-dimensional surface $\pd\Sigma$ \cite{Wald:1999wa},
\begin{equation}
    \label{eq-ham-2s}
    \slashed\delta Q_\xi[\pd\Sigma]=\oint_{\pd\Sigma}\left\{\delta Q_{ab}[\xi]-\xi^c\theta_{cab}[\delta]\right\},
\end{equation}
where $Q_{ab}[\xi]$ is the Neother charge 2-form.
For Chern-Simons modified gravity, the Neother charge 2-form $Q_{ab}[\xi]$ is
\begin{equation}
    \label{eq-neo-2f}
    Q_{ab}[\xi]=\kappa\epsilon_{abcd}\nabla^c\xi^d-\vartheta R_{abcd}\nabla^c\xi^d-{}^*R^*_{abcd}\xi^c\nabla^d\vartheta-2\xi^cR_{c[a}\nabla_{b]}\vartheta-R_{abcd}\xi^c\nabla^d\vartheta,
\end{equation}
where ${}^*R^*_{abcd}=\epsilon_{abef}{}^*R^{ef}{}_{cd}/2$ is the double Hodge dual, the first term also appears in general relativity and Brans-Dicke theory \cite{Wald:1999wa,Hou:2020wbo}, and the remaining terms come from $u^a[\delta_\xi]$.
To obtain this result, one has to show that ${}^*R^{cdea}(R_{cdeb}+2R_{cedb})=-{}^*R^{cdef}R_{dcef}\delta^a_b/2$ with Newman-Penrose variables \cite{Newman:1961qr}.
Note the symbol $\slashed\delta$ is used in eqs.~\eqref{eq-def-ham} and \eqref{eq-ham-2s}, because the right-hand sides might not be integrable.
As shown in ref.~\cite{Wald:1999wa}, the right-hand sides of eqs.~\eqref{eq-def-ham} and \eqref{eq-ham-2s} are integrable if and only if 
\begin{equation}
    \label{eq-iff}
    \oint_{\pd\Sigma}\xi^c\omega_{cab}[\delta,\delta']=0,
\end{equation}
provided that the exact and the linearized  equations of motion are satisfied.
This condition may not hold for some of the charges associated with the BMS generators with $\pd\Sigma$  a cross section of $\scri$. 
In order to define conserved charges for these BMS generators, one needs to modify eq.~\eqref{eq-ham-2s} by adding to it a correction according to ref.~\cite{Wald:1999wa}.
So now, let us explicitly carry out the calculation, and find the correction.

The calculation above is done in the physical spacetime. 
One wants to translate these results using unphysical quantities.
In the previous section, the conformal transformations of various quantities have been discussed. 
Here, one needs the transformations of the variations of these quantities.
Firstly, $\delta\Omega=0$, since there are no physical degrees of freedom associated with it.
Secondly, $\scri$ is a universal structure of any asymptotically flat spacetime \cite{Geroch1977}, so the unphysical metric $\hg_{ab}$ remains unchanged on $\scri$, that is,
\begin{equation}
    \label{eq-def-var-hg}
    \delta\hg_{ab}=\Omega\tau_{ab},\quad \delta g_{ab}=\Omega^{-1}\tau_{ab},
\end{equation}
for some smooth tensor field $\tau_{ab}$.
There is no special requirement on $\delta\hth(\equiv\chi)$.
Since Bondi gauge condition is very convenient, it should be preserved under the field variation, and this leads to 
\begin{equation}
    \label{eq-prop-tau}
    \tau_{ab}\hn^b=\Omega\tau_a,
\end{equation}
for yet another smooth vector field $\tau_a$.
Note that the variation \eqref{eq-def-var-hg} induces the variation in the Christoffel symbol,
\begin{equation}
    \label{eq-var-con}
    \delta\hat\Gamma^c{}_{ab}=\frac{1}{2}\hg^{cd}(\hn_a\tau_{bd}+\hn_b\tau_{ad}-\hn_d\tau_{ab})+\order{\Omega},
\end{equation}
whose restriction to $\scri$ determines $\delta\Sigma_{ab}$; see eq.~\eqref{eq-def-sig-0}.
Therefore, 
\begin{equation}
    \label{eq-tau-sig}
    \tau_{ab}\ddot=2\delta\Sigma_{ab}.
\end{equation}
With these, eqs.~\eqref{eq-psp} and \eqref{eq-psc} can be written in terms of the unphysical quantities.
The presymplectic potential current is found to be
\begin{equation}
    \label{eq-psp-up}
    \begin{split}
        \theta_{abc}[\delta]=&\heps_{dabc} \left\{\Omega^{-1} \left[ \kappa(\hnabla_e\tau^{de}-\hnabla^d\tau-3\tau^d)-\mb \chi\hth\hn^d \right] -\mb \chi\hnabla^d\hth \right\}\\
        &+\frac{\ma}{6}\hth\hn_{[a}\tau_b^d\hnabla_{c]}\hn_d+\order{\Omega},
    \end{split}
\end{equation}
where $\tau=\hg^{ab}\tau_{ab}$, and the indices of $\tau_{ab}$ are raised by $\hg^{ab}$.
The second line is from the Chern-Simons coupling term, where the first term evaluates to zero at $\scri$ due to the Bondi gauge condition, so it is also of $\order{\Omega}$.
The first line contains a term with $\Omega^{-1}$. 
It seems divergent on $\scri$, but not. 
This can be proved with the similar method in ref.~\cite{Hou:2020wbo}, as the equations of motion \eqref{eq-eoms-up} differ from eqs.~(13), (14) and (15) in ref.~\cite{Hou:2020wbo} by higher order terms in $\Omega$, which do not affect the argument in that reference.
We can now proceed to computing the presymplectic current 3-form,
\begin{equation}
    \label{eq-psc-up}
    \omega_{abc}=\heps_{abc} \left( \frac{\kappa}{2}\tau'^{de}\delta N_{de}+\mb\chi'\delta N \right)+\frac{\ma}{6}\chi\hn_{[a}\tau'^d_b\hnabla_{c]}\hn_d-\langle\delta\leftrightarrow\delta'\rangle+\order{\Omega},
\end{equation}
where $\tau'_{ab}=\Omega^{-1}\delta'\hg_{ab}$, $\chi'=\delta'\hth$, and the term with $\ma$ is of $\order{\Omega}$ with the Bondi gauge condition imposed.
These expressions for $\theta_{abc}$ and $\omega_{abc}$ actually resemble eqs.~(66) and (71) in ref.~\cite{Hou:2020wbo}, ignoring the terms of $\order{\Omega}$.
Therefore, in order to make eq.~\eqref{eq-ham-2s} integrable, one should add the following correction to $\theta_{abc}$,
\begin{equation}
    \label{eq-cor-psc}
    \Theta_{abc}[\delta]=\heps_{abc} \left( \frac{\kappa}{2}\tau^{de} N_{de}+\mb \chi N \right),
\end{equation}
which is defined on $\scri$.
There might be other choices of $\Theta_{abc}$, but according to the requirements made in ref.~\cite{Wald:1999wa}, this is the unique one.

\subsection{Fluxes and conserved charges}

With these, one can calculate the flux associated with $\xi^a$ through a patch $\mathscr B$ (a subset of $\scri$ bounded by cross sections $\mathscr C_1$ and $\mathscr C_2$), given by 
\begin{equation}
    \label{eq-flux}
    F_\xi[\mathscr B]=\int_{\mathscr B}\Theta_{abc}[\delta_\xi]=\int_{\mathscr B}\heps_{abc} \left( \kappa N_{de}\delta_\xi\Sigma^{de} +\mb N\delta_\xi\hth \right).
\end{equation}
So substituting eqs.~\eqref{eq-lieth}, \eqref{eq-liesig} and \eqref{eq-xi-dec}, one has 
\begin{equation}
    \label{eq-flux-v}
    \begin{split}
    F_\xi[\mathscr B] 
    =\int_{\mathscr B}\heps_{abc} &  \left\{ \left(\alpha+\frac{u}{2}\psi\right) \left[ \kappa \left( -\mathscr D_d\mathscr D_eN^{de}+\frac{1}{2}N^{de}N_{de} \right)+\mb N^2 \right] \right.\\
    &\left.+\kappa \left( N^{de}\lie_Y\sigma_{de}+\frac{\psi}{2}N^{de}\sigma_{de} \right) +\mb \left( N\lie_Y\hth+\frac{\psi}{2}N\hth \right) \right\},
    \end{split}
\end{equation}
where integration by parts has been used.
The flux $F_\xi[\mathscr B]$ can be rewritten in a different form.
Note that $\delta_\xi\Sigma_{ab}$ in eq.~\eqref{eq-flux} is actually $X_{ab}$ in eq.~\eqref{eq-sec-xab}, as both are related to the variation of the Levi-Civita connection $\hat\Gamma^c{}_{ab}$. 
Therefore, one has 
\begin{equation}
    \label{eq-flux-v-2}
    F_\xi[\mathscr B]=\int_{\mathscr B}\heps_{abc} \left\{ N^{de}\left[(\lie_\xi\mathscr D_d-\mathscr D_d\lie_\xi)\ell_e+\ell_{d}\mathscr D_{e}\psi\right] +\mb N\left(\lie_\xi\hth+\frac{\psi}{2}\hth\right)\right\}.
\end{equation}
By the conservation law, this flux should be given by the change in the corresponding Noether charge $Q_\xi[\mathscr C]$,
\begin{equation}
    \label{eq-flux-c}
    F_\xi[\mathscr B]=-(Q_\xi[\mathscr C_2]-Q_\xi[\mathscr C_1]),
\end{equation}
if $\mathscr C_2$ is in the future of $\mathscr C_1$.
These charges will be determined below.

In order to calculate the Noether charge, one also needs calculate eq.~\eqref{eq-neo-2f} in terms of the unphysical tensors, given by
\begin{equation}
    \label{eq-noe-2f-1}
    Q_{ab}[\xi]=-\kappa\heps_{abcd}\hnabla^c(\Omega^{-2}\xi^d)+2\Omega^{-2}\hat\vartheta(2\hn^c\hL_{c[a}\xi_{b]}-3\xi^c\hL_{c[a}\hn_{b]})+\order{\Omega}.
\end{equation}
Similarly, the first term is the same as the one in general relativity \cite{Wald:1999wa} and Brans-Dicke theory \cite{Hou:2020wbo}. 
It seems divergent on $\scri$, but is finite, which can be verified following the arguments in ref.~\cite{Hou:2020wbo}.
The remaining terms come from the Chern-Simons coupling term.
Due to the definition \eqref{eq-def-hL}, the second term above is actually of $\order{\Omega}$, so it, together with the third one, can be ignored for computing the conserved charges on $\scri$.
Now, split the BMS generator $\xi^a=\xi_1^a+\xi_2^a$ in the following way \cite{Flanagan:2015pxa},
\begin{equation}
    \label{eq-spxi}
    \xi_1^a=\frac{u-u_0}{2}\psi\hn^a+Y^a,\quad\xi_2^a= \left( \alpha+\frac{u_0}{2}\psi \right)\hn^a,
\end{equation}
where $\xi_1^a$ is tangent to the cross section $\mathscr C_0$ at $u=u_0$, and $\xi_2^a$ is a pure supertranslation.
The conserved charge for $\xi_1^a$ is simply \cite{Wald:1999wa}
\begin{equation}
    \label{eq-lor-ch}
    Q_{\xi_1}[\mathscr C_0]=\oint_{\mathscr C_0}Q_{ab}[\xi_1],
\end{equation}
at $\mathscr C_0$.
The conserved charge for $\xi_2^a$ has the following variation 
\begin{equation}
    \label{eq-sup-ch-de}
    \delta Q_{\xi_2}[\mathscr C_0]=\oint_{\mathscr C_0} \left\{ Q_{ab}[\alpha\hn] -\alpha\hn^c\theta_{cab}+\alpha\hn^c\Theta_{cab}\right\}.
\end{equation}
However, this expression is difficult to be integrated. 
So one can use eq.~\eqref{eq-flux-c} to find $Q_{\xi_2}[\mathscr C_0]$ \cite{Ashtekar:1981bq}
\begin{equation}
    \label{eq-sup-ch}
    Q_{\xi_2}[\mathscr C_0]=\oint_{\mathscr C_0} \left\{ 2\kappa \left[ \frac{\alpha'}{4}\hK^{de}\ell_e+N^{e[f}\hn^{d]}(\alpha'\mathscr D_f\ell_e+\ell_{f}\mathscr D_e\alpha') \right]-\frac{2\mb}{3}\alpha'\hn^d\hth N  \right\} \ell_d\hn^c\heps_{cab},
\end{equation}
with $\alpha'=\alpha+u_0\psi/2$.
The total charge is just $Q_\xi[\mathscr C_0]=Q_{\xi_1}[\mathscr C_0]+Q_{\xi_2}[\mathscr C_0]$.

With the asymptotic solution \eqref{eq-met-sols}, one can calculate the fluxes and charges in Bondi-Sachs coordinates. 
The flux associated with the supertranslation is 
\begin{equation}
    \label{eq-flux-sup}
    F_\alpha=\int\ud u\ud^2x\sqrt{\gamma} \alpha \left\{\kappa \left( \mathscr D_A\mathscr D_BN^{AB}+\frac{1}{2}N_{AB}N^{AB} \right) +\mb N^2 \right\},
\end{equation}
with $\gamma=\det(\gamma_{AB})$.
The flux conjugate to the Lorentz generator is 
\begin{equation}
    \label{eq-flux-lor}
    \begin{split}
        F_Y=F_{\tilde\alpha}+ \frac{1}{2}\int\ud u\ud^2x\sqrt\gamma Y^A & \bigg\{\kappa \bigg[ \frac{1}{2}(N_B^C\mathscr D_Ac_C^B-c_B^C\mathscr D_AN_C^B) \\
        &-\mathscr D^B(N^C_Bc_{AC}-c_B^CN_{AC}) \bigg]+\mb(N\mathscr D_A\vartheta_1-\vartheta_1\mathscr D_AN)\bigg\},
    \end{split}
\end{equation}
where $\tilde\alpha=u\psi/2$, and integration by parts has been performed.
The total charge  is given by 
\begin{equation}
    \label{eq-neoc}
    Q_\xi[\mathscr C]=\int\ud^2\sqrt\gamma \left\{\kappa \left[ 4\alpha m-2u\lie_Ym+2Y^A\hat N_A+\frac{1}{16}\lie_Y(c_{A}^{B}c^{A}_{B})\right]+\frac{\mb}{4}\vartheta_1\lie_Y\vartheta_1  \right\},
\end{equation}
where $\hat N_A=N_A-\frac{3}{32}\mathscr D_A(c_B^Cc^B_C)-\frac{1}{4}c_{A}^{B}\mathscr D_Cc^{B}_{C}$.
This charge is defined for an arbitrary $u$.
More specifically, the supermomentum is given by the following integral
\begin{equation}
    \label{eq-def-supm}
    \mathcal P_\alpha[\mathscr C]=4\kappa\int\ud^2x\sqrt\gamma\alpha m.
\end{equation}
If $\alpha$ is a linear combination of $l=0,1$ spherical harmonics, $\mathcal P_\alpha$ is a linear combination of the ordinary momenta.
In particular, when $\alpha=1$, eq.~\eqref{eq-def-supm} is the Bondi mass, or the total energy of the gravitating system.
The supermomentum is a generalization of the ordinary momentum.
The charge conjugate to $Y^A$ can be catgorized to two groups. 
Decompose $Y^A$ in the following way,
\begin{equation}
    \label{eq-dec-y}
    Y^A=\mathscr D^A\mu+\hat\epsilon^{AB}\mathscr D_B\upsilon,
\end{equation}
where $\mu$ and $\upsilon$ are linear combinations of $l=1$ spherical harmonics with $(\mathscr D^2+2)\mu=(\mathscr D^2+2)\upsilon=0$. 
$\mu$ is called the electric part of $Y^A$, and generates the Lorentz boost, whose charge is 
\begin{equation}
    \label{eq-bst-c}
    \mathcal K_\mu[\mathscr C]=-\int\ud^2x\sqrt\gamma \mu \left[ 2\kappa \left(\mathscr D^A\hat N_A+2um-\frac{c_{AB}c^{AB}}{16}\right)-\frac{\mb}{4}\vartheta_1^2 \right].
\end{equation}
$\upsilon$ is the magnetic part of $Y^A$, and generates the spatial rotation with the following angular momentum
\begin{equation}
    \label{eq-ang-c}
    \mathcal J_\upsilon[\mathscr C]=-2\kappa\int\ud^2x\sqrt\gamma\upsilon\hat\epsilon^{AB}\mathscr D_A\hat N_B.
\end{equation}
$\mathcal K_\mu$ and $\mathcal J_\upsilon$ can also be called the CM and the spin charges, respectively.

\subsection{Constraints on memory effects}
\label{sec-cst-mm}

As discussed in ref.~\cite{Hou:2021cbd}, memory effects can be measured by letting two test particles freely fall and monitoring their relative distance. 
Let $d^a$ represent the spatial deviation vector between two test particles and tangent to $\scri$ when evaluated on it, then one has the following geodesic deviation equation \cite{Misner:1974qy},
\begin{equation}
    \label{eq-gde}
    \hn^c\nabla_c(\hn^b\nabla_bd^a)=-R_{cbd}{}^a\hn^cd^b\hn^d,
\end{equation}
evaluated near $\scri$, where $\hn^a=(\pd_u)^a$ is approximately the 4-velocity of the test particle \cite{Strominger:2014pwa}.
Taking the conformal transformation of the right-hand side, one has 
\begin{equation}
    \label{eq-gde-cft}
    \ddot d^a=\frac{\Omega}{4}\hK^a{}_bd^b+\order{\Omega^2}.
\end{equation}
Contracting eq.~\eqref{eq-shk-div} with $\hn^a$ and lowering the index $c$, one gets 
\begin{equation}
    \label{eq-hknews}
    \hK_{ab}=2\lie_{\hn}N_{ab}\equiv2\dot N_{ab}.
\end{equation}
Therefore, the geodesic deviation equation becomes 
\begin{equation}
    \label{eq-gde-v}
    \ddot d^a=\frac{\Omega }{2}d^b\dot N^{a}_{b}+\order{\Omega^2}.
\end{equation}
Integrating this equation, one gets the total changes in $\dot d^a$ and $d^a$,
\begin{subequations}
    \begin{gather}
        \Delta\dot d^a=-\Omega d^b\Delta \sigma^{a}_{b}+\order{\Omega^2},\\
        \Delta d^a=\Omega \{d^b\Delta\sigma^{a}_{b}+\dot d^b[(\sigma^{a}_{b}|_f+\sigma^{a}_{b}|_0)\Delta u-\Delta\mathcal C^{a}_{b}]\}+\order{\Omega^2},\label{eq-dism}
    \end{gather}
\end{subequations}
where $\sigma^{a}_{b}|_f$ means to evaluate $\sigma^{a}_{b}$ at the retarded time $u=u_f$, similarly defined for $\sigma^{a}_{b}|_0$, $\Delta u=u_f-u_0$, $\Delta\sigma^{a}_{b}=\sigma^{a}_{b}|_f-\sigma^{a}_{b}|_0$, and $\Delta\mathcal C^{a}_{b}=2\int_{u_0}^{u_f}\sigma^{a}_{b}\ud u$.
If one assumes that the spacetime is nonradiative before $u_0$ and after $u_f$, then $\Delta\dot d^a$ represents the velocity kick memory \cite{Seraj:2021qja}, and $\Delta d^a$ the displacement memory. 
In particular, the first term in the curly brackets in eq.~\eqref{eq-dism} is the leading displacement memory, and the remaining terms represent the subleading displacement memory.
Usually, the leading displacement memory is also called displacement memory, as mentioned before.
Spin memory and CM memory contribute to the subleading displacement memory as discussed below.
Using quantities defined in section~\ref{sec-cs}, one recovers eqs.~(61) and (62) in ref.~\cite{Hou:2021cbd}. 
For reference, we copy them here,
\begin{subequations}
    \begin{gather}
        \Delta\dot d_{\hat A}\approx-\frac{\Delta c_{AB}}{2r}\dot d_0^{\hat B},\label{eq-def-vk}\\
        \Delta d_{\hat A}\approx\dot d_{\hat A}^0\Delta u+\frac{\Delta c_{AB}}{2r}d^{\hat B}_0
        +\frac{1}{r}\left[\frac{c_{AB}(u_f)+c_{AB}(u_0)}{2}\Delta u-\Delta\mathcal C_{AB}\right]\dot d_0^{\hat B},\label{eq-def-3mm}
    \end{gather}
\end{subequations}
where $\hat A,\hat B$ correspond to the normalized basis in the angular directions.
In the following, we will work in the Bondi-Sachs coordinates.

To get the constraint on the (leading) displacement memory and the velocity kick memory, one uses the flux-balance law for an arbitrary supertranslation $\alpha$,
\begin{equation}
    \label{eq-fb-sup}
    F_\alpha=-\Delta \mathcal P_\alpha=-[\mathcal P(u_f)-\mathcal P(u_0)].
\end{equation}
Let
\begin{equation}
    \label{eq-dec-c}
c_{AB}=\left(\mathscr D_A\mathscr D_B-\frac{1}{2}\gamma_{AB}\mathscr D^2\right)\Phi+\hat\epsilon_{C(A}\mathscr D_{B)}\mathscr D^C\Psi,
\end{equation}
for arbitrary functions $\Phi$ and $\Psi$.
Then, substituting eq.~\eqref{eq-flux-sup} in and integrating by parts, one gets 
\begin{equation}
    \label{eq-cst-dism}
    \int\ud^2x\sqrt\gamma \alpha\mathscr D^2(\mathscr D^2+2)\Delta\Phi=\frac{2}{\kappa}(\mathcal E_\alpha+\Delta\mathcal P_\alpha),
\end{equation}
where $\Delta\Phi=\Phi(u_f)-\Phi(u_0)$, and $\mathcal E_\alpha$ is eq.~\eqref{eq-flux-sup} without the term linear in $N_{AB}$ in the integrand.
Since in vacuum, eq.~\eqref{eq-def-vac} holds, $\Delta\Phi$ measures the displacement and velocity kick memories.
The spin memory is determined by \cite{Pasterski:2015tva,Flanagan:2015pxa}
\begin{equation}
    \label{eq-def-sm}
    \Delta\mathcal S=\int_{u_0}^{u_f}\ud u\Psi.
\end{equation}
To determine the constraint on this memory, one needs the flux-balance law associated with the spatial rotation,
\begin{equation}
    \label{eq-fb-rot}
    F_\upsilon=-\Delta\mathcal  J_\upsilon=-[\mathcal J_\upsilon(u_f)-\mathcal J_\upsilon(u_0)],
\end{equation}
where the left-hand side is eq.~\eqref{eq-flux-lor} with $Y^A=\hat\epsilon^{AB}\mathscr D_B\upsilon$.
Direct calculation shows that this does not lead to any constraint, because $(\mathcal D^2+2)\upsilon=0$.
To resolve this problem, one wants to extend BMS group by allowing $\upsilon$ be more general functions on the unit 2-sphere, probably singular at finite points. 
But this may require us to determine new fluxes and conserved charges, which can be tricky due to the singularities.
Instead, following ref.~\cite{Flanagan:2015pxa}, we simply modify the flux \eqref{eq-flux-lor} by adding the following correction\footnote{The basic idea to obtain the correction $\mathscr F_Y$ is to change the flux \eqref{eq-flux-lor} such that for any $Y^A$ in the Virasoro algebra, the flux-balance law $F'_Y[\mathscr B]=-(Q_Y[\mathscr C_2]-Q_Y[\mathscr C_1])$ still holds.
Note that we want to keep the definition of $Q_Y$ unchanged.
In order to obtain the correction \eqref{eq-flux-lor-c}, one contracts eq.~\eqref{eq-evo-n} by $2\kappa Y^A$, and integrates the both sides over the null infinity. 
One finds out that the first round brackets on the right-hand side of eq.~\eqref{eq-evo-n} gives the integrand in eq.~\eqref{eq-flux-lor-c}, and the remaining terms in eq.~\eqref{eq-evo-n} can be rearranged to reproduce the correct charge and the original flux for the global conformal Killing vector field $Y^A$.
In the actual computation, one explores the integration by parts, and eq.~\eqref{eq-evo-m}.}
\begin{equation}
    \label{eq-flux-lor-c}
    \begin{split}
    \mathscr F_Y=&\frac{\kappa}{2}\int\ud u\ud^2x\sqrt{\gamma} Y^A\mathscr D^B(\mathscr D_A\mathscr D_Cc_B^C-\mathscr D_B\mathscr D_Cc_A^C)\\
        &=\frac{\kappa}{4}\int\ud u\ud^2x\sqrt\gamma \hat\epsilon_{AB}Y^A\mathscr D^B\mathscr D^2(\mathscr D^2+2)\Psi.
    \end{split}
\end{equation}
In this way, the conserved charges remain the same, and the flux-balance law eq.~\eqref{eq-fb-rot} now takes the form $F_\upsilon+\mathscr F_{\upsilon}=-[\mathcal J_\upsilon(u_f)-\mathcal J_\upsilon(u_0)]$.
Then, by rearranging the modified flux-balance law, one obtains the following constraints on spin memory,
\begin{equation}
    \label{eq-cst-sm}
    \int\ud^2x\sqrt\gamma\upsilon\mathscr D^2\mathscr D^2(\mathscr D^2+2)\Delta\mathcal S=\frac{4}{\kappa}(\Delta\mathcal J_\upsilon+\mathcal Q_\upsilon+\mathscr J_\upsilon),
\end{equation}
with 
\begin{gather}
    \label{eq-defs}
    \mathscr J_\upsilon=\int\ud u\ud^2x\sqrt\gamma\hat\epsilon^{AB}\mathscr D_AJ_B,\quad J_A=\frac{\kappa}{2}N_B^C\mathscr D_Ac_C^B+\mb N\mathscr D_A\vartheta_1,\\
    \mathcal Q_\upsilon=-\frac{\kappa}{2}\int\ud u\ud^2x\sqrt{\gamma}\upsilon\mathscr D^2(\hat\epsilon^{AB}N_{AC}c_B^C).
\end{gather}
Finally, let us work out the constraint on CM memory \cite{Nichols:2018qac}.
So one writes $\Phi=\Phi_n+\Phi_o$ such that 
\begin{equation}
    \int\ud^2x\sqrt\gamma\alpha\mathscr D^2(\mathscr D^2+2)\Delta\Phi_n=\frac{2}{\kappa}\mathcal E_\alpha,\quad \int\ud^2x\sqrt\gamma\alpha\mathscr D^2(\mathscr D^2+2)\Delta\Phi_o=\frac{2}{\kappa}\Delta\mathcal P_\alpha.
\end{equation}
Then CM memory is measured by the following integral,
\begin{equation}
    \label{eq-def-cm}
    \Delta\mathcal K=\int_{u_0}^{u_f}u\dot\Phi_o\ud u.
\end{equation}
It satisfies the flux-balance law  associated with the superboost,
\begin{equation}
    \label{eq-fb-sb}
    F_\mu=-\Delta\mathcal K_\mu=-[\mathcal K_\mu(u_f)-\mathcal K_\mu(u_0)],
\end{equation}
where $F_\mu$ is given by eq.~\eqref{eq-flux-lor} with $Y^A=\mathscr D^A\mu$.
It turns out that 
\begin{equation}
    \label{eq-sct-cm}
    \int\ud^2x\sqrt\gamma\mu\mathscr D^2\mathscr D^2(\mathscr D^2+2)\Delta\mathcal K=4\kappa(\mathscr J'_\mu+\mathcal Q'_\mu-\Delta\mathcal K_\mu),
\end{equation}
with 
\begin{equation}
    \mathscr J'_\mu=\int\ud u\ud^2x\sqrt{\gamma}\mu\mathscr D^AJ_A, \quad \mathcal Q'_\mu=-\frac{1}{2}\int\ud u\ud^2x\sqrt{\gamma}\mu\mathscr D^2\left(\frac{\kappa}{2}c_A^BN_B^A+\mb\vartheta_1N\right).
\end{equation}
Like $\upsilon$ in eq.~\eqref{eq-cst-sm}, $\mu$ can have a finite number of singularities on the unit 2-sphere.

\section{Chern-Simons modified gravity in first-order formalism}
\label{sec-fir}

In the previous work \cite{Hou:2021cbd}, it was also assumed that $\vartheta$ does not interact with the ordinary matter fields directly, so it is impossible to use laser interferometers, pulsar timing arrays or the Gaia mission to detect the scalar gravitational wave, not to mention its memory effect, even if it exists.
Therefore, the scalar memory was not discussed.
Here, the memory effect in the scalar sector will be studied. 
For this purpose, one uses the idea proposed in refs.~\cite{Campiglia:2017dpg,Campiglia:2018see}.
There, a free scalar field theory is equivalent to a gauge theory with the gauge field being a differential 2-form.
Then, the new theory possesses the gauge symmetry, which is nontrivial at the null infinity, and is related to the memory effect of the scalar field.
This treatment has also been used to explain the S memory in Brans-Dicke theory in ref.~\cite{Seraj:2021qja}.

So now, for Chern-Simons modified gravity, one should also find an equivalent description, where a 2-form field plays the role of Chern-Simons scalar field.
It turns out that there already exists such kind of formulation in ref.~\cite{Yoshida:2019dxu}, and the action is more naturally written with differential forms.
Thus, we will follow this method to rewrite the action~\eqref{eq-cs-act} in the first-order formalism.
In the first-order formalism, the tetrad and the connection are independent of each other, and in general, torsion is allowed.
Let $\{\bvec{e}{\mu}{a}\}$ be tetrad basis, and its dual is $\{\bform{\theta}{\mu}{a}\}$.
In this basis, the metric $g_{ab}$ has the components $g_{\hat\mu\hat\nu}$, which and whose inverse $g^{\hat\mu\hat\nu}$ are used to raise and lower the hatted indices (or, internal indices as in gauge theories of gravity \cite{Blagojevic:2002du,Aldrovandi:2013bk,Fabi:2013bik}).
Here, for simplicity, one requires $g_{\hat\mu\hat\nu}$ be constant. 
It can be $\eta_{\hat\mu\hat\nu}=\text{diag}\{-1,1,1,1\}$ or any form that is convenient.
Cartan's first and second structure equations define the torsion and curvature 2-forms \cite{Blagojevic:2002du}
\begin{gather}
    \label{eq-def-tor}
    \torsion{\mu}{ab}=\ud_a\bform{\theta}{\mu}{b}-\scon{\nu}{\mu}{a}\wedge\bform{\theta}{\nu}{b},\\
    \label{eq-def-cur}
    \riem{ab}{\mu}{\nu}=\ud_a\scon{\mu}{\nu}{b}+\scon{\mu}{\rho}{a}\wedge\scon{\rho}{\nu}{b}\equiv-\frac{1}{2}\mathcal R^{\hat\nu}{}_{\hat\mu\hat\rho\hat\sigma}\bform{\theta}{\rho}{a}\wedge\bform{\theta}{\sigma}{b},
\end{gather} 
respectively.
Here, $\scon{\nu}{\mu}{a}$ is the connection 1-form, $\mathcal R^{\hat\nu}{}_{\hat\mu\hat\rho\hat\sigma}$ is the curvature tensor $\mathcal R^d{}_{cab}$ in the tetrad basis, which is not generally the Riemann tensor $R^{\hat\nu}{}_{\hat\mu\hat\rho\hat\sigma}$ introduced in action~\eqref{eq-cs-act}.
One can show that 
\begin{gather}
    \Sigma^{\hat\mu\hat\nu}_{ab}\wedge\Omega_{cd\hat\mu\hat\nu}=\mathcal R\epsilon_{abcd},\quad
    \riem{ab}{\mu}{\nu}\wedge\riem{cd}{\nu}{\mu}=-\frac{1}{2}{}^*\mathcal R_{\hat\mu\hat\nu\hat\rho\hat\sigma}\mathcal R^{\hat\nu\hat\mu\hat\rho\hat\sigma}\epsilon_{abcd},
\end{gather}
where $\Sigma^{\hat\mu\hat\nu}_{ab}=\frac{1}{2}\epsilon^{\hat\mu\hat\nu}{}_{\hat\rho\hat\sigma}\bform{\theta}{\rho}{a}\wedge\bform{\theta}{\sigma}{b}$ and $\epsilon^{\hat\mu\hat\nu}{}_{\hat\rho\hat\sigma}$ represents the component of the volume element $\epsilon^{ab}{}_{cd}$ in the chosen basis.
The second equation above defines the Pontryagin term \cite{Corichi:2013zza}.
Using eq.~\eqref{eq-def-cur}, one can find out that \cite{Yoshida:2019dxu},
\begin{gather}
    \riem{ab}{\mu}{\nu}\wedge\riem{cd}{\nu}{\mu}=\ud_a{}^*J_{bcd},\quad 
    {}^*J_{abc}=\scon{\mu}{\nu}{a}\wedge\ud_b\scon{\nu}{\mu}{c}+\frac{2}{3}\scon{\mu}{\rho}{a}\wedge\scon{\rho}{\nu}{b}\wedge\scon{\nu}{\mu}{c},
\end{gather}
where ${}^*J_{abc}$ is also called the gravitational Chern-Simons anomalous term in string theory \cite{Campbell:1990fu}.
These will be useful for transforming the action~\eqref{eq-cs-act}.
In the following, the abstract indices will be omitted and the kernel letter  will be in bold face for each differential form.
The abstract indices are sometimes written explicitly for the sake of clearance.

The action~\eqref{eq-cs-act} can be rewritten in the following form,
\begin{equation}
    S=\int\left[ \kappa \boldsymbol\Sigma^{\hat\mu\hat\nu}\wedge\boldsymbol{\Omega}_{\hat\mu\hat\nu}+\frac{\ma}{2}\ud\vartheta\wedge {}^*\boldsymbol J-\frac{\mb}{2}\ud\vartheta\wedge {}^*(\ud\vartheta) \right],
\end{equation}
where integration by part has been used.
According to ref.~\cite{Yoshida:2019dxu}, in order to obtain the dual 2-form equivalent action, one first introduces two new forms $A_a$ and $B_{ab}$ such that 
\begin{equation}
    \label{eq-new-act}
    S_1=\int\left[ \kappa \boldsymbol\Sigma^{\hat\mu\hat\nu}\wedge\boldsymbol{\Omega}_{\hat\mu\hat\nu}+\frac{\ma}{2}\boldsymbol A\wedge {}^*\boldsymbol J-\frac{\mb}{2}\boldsymbol A\wedge {}^*\boldsymbol A+\frac{1}{2}\boldsymbol B\wedge\ud\boldsymbol A\right].
\end{equation}
The variation of this new action with respect to $\boldsymbol B$ leads to $\ud\boldsymbol A=0$, so locally $\boldsymbol A$ is the differential of a scalar field, say $\boldsymbol A=\ud\vartheta$.
The variation with respect to $\boldsymbol A$ gives rise to 
\begin{equation}
    \label{eq-eom-a}
    ^*\boldsymbol A=\frac{\boldsymbol{\hat H}}{2\mb},\quad \boldsymbol{\hat H}=\ud\boldsymbol B+\ma {}^*\boldsymbol J.
\end{equation}
Now, plug these relations back into the action~\eqref{eq-new-act}, then one obtains,
\begin{equation}
    \label{eq-act-eq}
    S_2=\int\left[ \kappa \boldsymbol\Sigma^{\hat\mu\hat\nu}\wedge\boldsymbol{\Omega}_{\hat\mu\hat\nu}-\frac{1}{8\mb}{}^*\boldsymbol{\hat H}\wedge\boldsymbol{\hat H}\right].
\end{equation}
But this action leads to a nontrivial torsion \cite{Alexander:2008wi,BottaCantcheff:2008pii}, so let us add a  Lagrange multiplier $\mathfrak L^{\hat\mu}{}_{ab}$ which is antisymmetric in exchanging the two lower indices, and the final action is
\begin{equation}
    \label{eq-act-eq-cs}
    S_3=\int\left[ \kappa \boldsymbol\Sigma^{\hat\mu\hat\nu}\wedge\boldsymbol{\Omega}_{\hat\mu\hat\nu}-\frac{1}{8\mb}{}^*\boldsymbol{\hat H}\wedge\boldsymbol{\hat H}+\boldsymbol{\mathcal T}^{\hat\mu}\wedge\boldsymbol{\mathfrak L}_{\hat\mu}\right].
\end{equation}
This will be viewed as a fundamental action.
If one ignores the term with $\boldsymbol{\mathfrak L}_{\hat\mu}$ and forces the torsion to vanish by hand, $S_3$ resembles the effective 4-dimensional string action for heterotic and type II string theory with Gauss-Bonnet terms, dilaton terms and matter terms omitted \cite{Campbell:1990fu,Campbell:1992hc}. 
In this effective action, $\boldsymbol{\hat H}$ is called the Kalb-Ramond 3-form field strength, and ${\boldsymbol\omega}_{\hat\mu}{}^{\hat\nu}$ is the Lorentz-Chern-Simons term.
The equations of motion has been calculated for the effective action in ref.~\cite{Smith:2007jm}, and it turns out that they are equivalent to eqs.~\eqref{eq-eoms} once one identifies ${}^*\boldsymbol{\hat H}/2\mb=\ud\vartheta$ with $\vartheta$ called the Kalb-Ramond axion or the universal axion.

Before deriving equations of motion, let us discuss the symmetries of $S_3$.
It is not only diffeomorphism invariant, but gauge invariant under the following transformation,
\begin{equation}
    \label{eq-gtf}
    \boldsymbol B\rightarrow\boldsymbol B+\ud\boldsymbol{\mathfrak A},
\end{equation}
for an arbitrary 1-form field $\boldsymbol{\mathfrak A}$.
In fact, it also possesses the local Lorentz invariance.
If the tetrad basis is chosen such that $g_{\hat\mu\hat\nu}=\eta_{\hat\mu\hat\nu}$, the local Lorentz transformation parameterized by $\Lambda^{\hat\mu}{}_{\hat\nu}(x)$ transforms various tensors in the following way \cite{Blagojevic:2002du},
\begin{equation}
    \label{eq-llt}
    \eta_{\hat\mu\hat\nu}=\Lambda^{\hat\rho}{}_{\hat\mu}\Lambda^{\hat\sigma}{}_{\hat\nu}\eta_{\hat\rho\hat\sigma},\quad\boldsymbol\theta'^{\hat\mu}=\Lambda^{\hat\mu}{}_{\hat\nu}\boldsymbol\theta^{\hat\nu},\quad\boldsymbol\omega'_{\hat\mu}{}^{\hat\nu}=\Lambda_{\hat\mu}{}^{\hat\rho}\boldsymbol\omega_{\hat\rho}{}^{\hat\sigma}\Lambda^{\hat\nu}{}_{\hat\sigma}+\Lambda_{\hat\mu}{}^{\hat\rho}\ud\Lambda^{\hat\nu}{}_{\hat\rho},
\end{equation}
where $\Lambda_{\hat\mu}{}^{\hat\nu}$ is the inverse Lorentz transformation satisfying $\Lambda_{\hat\mu}{}^{\hat\rho}\Lambda^{\hat\nu}{}_{\hat\rho}=\delta^{\hat\nu}_{\hat\mu}$.
The first expression is the definition of the local Lorentz transformation, the second means that $\boldsymbol\theta^{\hat\mu}$ is a local Lorentz vector, and the third states that $\boldsymbol\omega_{\hat\mu}{}^{\hat\nu}$ is the gauge potential of the transformation.
Under the local Lorentz transformation, both $\boldsymbol\Omega_{\hat\mu}{}^{\hat\nu}$ and $\boldsymbol{\mathcal T}^{\hat\mu}$ transform covariantly, 
\begin{equation}
    \label{eq-def-tct}
    \boldsymbol\Omega'_{\hat\mu}{}^{\hat\nu}=\Lambda_{\hat\mu}{}^{\hat\rho}\boldsymbol\Omega_{\hat\rho}{}^{\hat\sigma}\Lambda^{\hat\nu}{}_{\hat\sigma},\quad \boldsymbol{\mathcal T}'^{\hat\mu}=\Lambda^{\hat\mu}{}_{\hat\nu}\boldsymbol{\mathcal T}^{\hat\nu},
\end{equation}
so $\boldsymbol{\mathfrak L}^{\hat\mu}\rightarrow\Lambda^{\hat\mu}{}_{\hat\nu}\boldsymbol{\mathfrak L}^{\hat\nu}$.
But ${}^*\boldsymbol J$ transforms according to 
\begin{equation}
    \label{eq-tsj}
    {}^*\boldsymbol J'={}^*\boldsymbol J+\ud\boldsymbol\omega_{\hat\mu}{}^{\hat\nu}\wedge\boldsymbol\Delta_{\hat\nu}{}^{\hat\mu}-\boldsymbol\omega_{\hat\mu}{}^{\hat\nu}\wedge\boldsymbol\Delta_{\hat\nu}{}^{\hat\rho}\wedge\boldsymbol\Delta_{\hat\rho}{}^{\hat\mu}-\frac{1}{3}\boldsymbol\Delta_{\hat\mu}{}^{\hat\nu}\wedge\boldsymbol\Delta_{\hat\nu}{}^{\hat\rho}\wedge\boldsymbol\Delta_{\hat\rho}{}^{\hat\mu},
\end{equation}
with $\boldsymbol\Delta_{\hat\mu}{}^{\hat\nu}=\Lambda_{\hat\rho}{}^{\hat\nu}\ud\Lambda^{\hat\rho}{}_{\hat\mu}$.
So ${}^*\boldsymbol J$ is not invariant. 
To make sure the action $S_3$ is local Lorentz invariant, one requires that $\boldsymbol{\hat H}$ be invariant, and $\boldsymbol B$ transform to compensate the change in ${}^*\boldsymbol J$.
It is difficult to write the finite transformation law of $\boldsymbol B$, but if $\lambda^{\hat\mu}{}_{\hat\nu}$ represents an infinitesimal local Lorentz transformation, such that $\Lambda^{\hat\mu}{}_{\hat\nu}=\delta^{\hat\mu}_{\hat\nu}+\lambda^{\hat\mu}{}_{\hat\nu}$ and $\Lambda_{\hat\nu}{}^{\hat\mu}=\delta^{\hat\mu}_{\hat\nu}-\lambda^{\hat\mu}{}_{\hat\nu}$, one knows that 
\begin{equation}
        \delta_\lambda\boldsymbol\theta^{\hat\mu}=\lambda^{\hat\mu}{}_{\hat\nu}\boldsymbol\theta^{\hat\nu},\quad \delta_\lambda\boldsymbol\omega_{\hat\mu}{}^{\hat\nu}=\mathcal D\lambda^{\hat\nu}{}_{\hat\mu},\quad
        \delta_\lambda{}^*\boldsymbol J=\ud\boldsymbol\omega_{\hat\mu}{}^{\hat\nu}\wedge\ud\lambda^{\hat\mu}{}_{\hat\nu},\quad
        \delta_\lambda\boldsymbol B=-\ma\boldsymbol\omega_{\hat\mu}{}^{\hat\nu}\wedge\ud\lambda^{\hat\mu}{}_{\hat\nu}.
\end{equation}
Here, one also defines the ``covariant exterior derivative'' $\mathcal D$, acting on an arbitrary differential form $\boldsymbol\lambda^{\hat\mu\cdots\hat\nu}_{\hat\rho\cdots\hat\sigma}$ in the following way,
\begin{equation}
    \label{eq-def-md}
    \mathcal D\boldsymbol\lambda^{\hat\mu\cdots\hat\nu}_{\hat\rho\cdots\hat\sigma}=\ud\boldsymbol\lambda^{\hat\mu\cdots\hat\nu}_{\hat\rho\cdots\hat\sigma}+\boldsymbol\omega_{\hat\rho}{}^{\hat\alpha}\wedge\boldsymbol\lambda^{\hat\mu\cdots\hat\nu}_{\hat\alpha\cdots\hat\sigma}+\cdots+\boldsymbol\omega_{\hat\sigma}{}^{\hat\alpha}\wedge\boldsymbol\lambda^{\hat\mu\cdots\hat\nu}_{\hat\rho\cdots\hat\alpha}-\boldsymbol\omega_{\hat\alpha}{}^{\hat\mu}\wedge\boldsymbol\lambda^{\hat\alpha\cdots\hat\nu}_{\hat\rho\cdots\hat\sigma}-\cdots-\boldsymbol\omega_{\hat\alpha}{}^{\hat\nu}\wedge\boldsymbol\lambda^{\hat\mu\cdots\hat\alpha}_{\hat\rho\cdots\hat\sigma}.
\end{equation}
In fact, $\boldsymbol{\mathcal T}^{\hat\mu}=\mathcal D\boldsymbol\theta^{\hat\mu}$.

Now, one can determine the equations of motion, given by
\begin{subequations}
    \label{eq-eoms-fof}
    \begin{gather}
        \kappa\epsilon^{\hat\rho\hat\sigma}{}_{\hat\nu\hat\mu}\boldsymbol{\Omega}_{\hat\rho\hat\sigma}\wedge\boldsymbol\theta^{\hat\nu}-\mathcal D\boldsymbol{\mathfrak L}_{\hat\mu}+\frac{1}{4\mb}{}^*\boldsymbol{\hat H}\wedge\boldsymbol{\hat H}_{\hat\mu}-\frac{1}{8\mb}{}^*({}^*\boldsymbol{\hat H}\wedge\boldsymbol{\hat H}){}^*\boldsymbol\theta_{\hat\mu}=0, \label{eq-eom-1f}\\
        \kappa \mathcal D\boldsymbol\Sigma^{\hat\mu\hat\nu}-\boldsymbol\theta^{[\hat\mu}\wedge\boldsymbol{\mathfrak L}^{\hat\nu]}-\frac{\ma}{2\mb}{}^*\boldsymbol{\hat H}\wedge\boldsymbol{\Omega}^{\hat\mu\hat\nu}=0,\label{eq-eom-sc}\\
        \ud{}^*\boldsymbol{\hat H}=0,\label{eq-eom-b}\\
        \boldsymbol{\mathcal T}^{\hat\mu}=0,
    \end{gather}
\end{subequations}
where $\hat H_{\hat\mu ab}=\bvec{e}{\mu}{c}\hat H_{cab}$ is the interior product.
Equation~\eqref{eq-eom-b} allows one to identify $\ud \vartheta\propto{}^*\boldsymbol{\hat H}$, at least locally \cite{Campbell:1990fu,Smith:2007jm,Yoshida:2019dxu}.

We can rewrite eqs.~\eqref{eq-eoms-fof} to make them similar to eqs.~\eqref{eq-eoms}.
Now, start with eq.~\eqref{eq-eom-sc}.  
Take the Hodge dual, and then after some algebraic manipulation, one has
\begin{equation}
    \label{eq-eom-sc-1}
    {}^*\mathfrak L^{bca}-{}^*\mathfrak L^{cba}=-\frac{\ma}{\mb}{}^*\hat H_d{}^*R^{bcda},
\end{equation}
where ${}^*\mathfrak L^c{}_{ab}=(e_{\hat\mu})^c\mathfrak L^{\hat\mu}{}_{de}\epsilon^{de}{}_{ab}/2$.
Since ${}^*\mathfrak L^{cab}={}^*\mathfrak L^{c[ab]}$, one gets 
\begin{equation}
    \label{eq-lfm}
    {}^*\mathfrak L^{cab}=-\frac{\ma}{2\mb}{}^*\hat H_{d}({}^*R^{cabd}+{}^*R^{bcad}+{}^*R^{abcd}).
\end{equation}
Then, take the Hodge dual of eq.~\eqref{eq-eom-1f} using the above result to get,
\begin{equation}
    \label{eq-eom-1f-1}
   \kappa[2R_{\hat\mu}{}^a-\bvec{e}{\mu}{a}R]+2C^a_{\hat\mu}=\frac{1}{4\mb}\left( {}^*\hat H_{\hat\mu}{}^*\hat H^a-\frac{1}{2}\bvec{e}{\mu}{a}{}^*\hat H_b{}^*\hat H^b \right),
\end{equation}
where the C-tensor $C^a_{\hat\mu}$ is given by
\begin{equation}
    \label{eq-def-dots}
    \begin{split}
    C^a_{\hat\mu}=&\frac{1}{12}\epsilon^{abcd}(\ud_b\mathfrak L_{\hat\mu cd}+\scon{\mu}{\nu}{b}\wedge\mathfrak L_{\hat\nu cd})\\
    =&-\frac{\ma}{2\mb}g_{\hat\mu\hat\nu}\bvec{e}{\rho}{a}({}^*\hat H_{\hat\sigma}\epsilon^{\hat\sigma\hat\alpha\hat\beta(\hat\rho}\nabla_{\hat\beta}R_{\hat\alpha}^{\hat\nu)}+{}^*R^{\hat\lambda(\hat\rho\hat\nu)\hat\sigma}\nabla_{\hat\sigma}{}^*\hat H_{\hat\lambda}).
    \end{split}
\end{equation}
To obtain the above relations, one uses the following results,
\begin{equation}
    \nabla_b\bvec{e}{\mu}{a}=-\bvec{e}{\nu}{a}\scon{\mu}{\nu}{b}, \quad \nabla_{\hat\nu}u^{\hat\mu}=\pd_{\hat\nu}u^{\hat\mu}-\omega_{\hat\rho}{}^{\hat\mu}{}_{\hat\nu}u^{\hat\rho},\quad \nabla_{\hat\nu}w_{\hat\mu}=\pd_{\nu} w_{\hat\mu}+\omega_{\hat\mu}{}^{\hat\rho}{}_{\hat\nu}w_{\hat\rho},
\end{equation}
for any vector field $u^a$ and dual vector field $w_a$,
and of course, the Bianchi identities.
Eventually, Einstein's equation~\eqref{eq-ein} can be recovered by contracting eq.~\eqref{eq-eom-1f-1} by $\bform{\theta}{\mu}{c}g_{ab}/2\kappa$ and identifying $\nabla_a\vartheta={}^*\hat H_a/2\mb$.
Finally, the second expression in eq.~\eqref{eq-eom-a} can be explicitly written as
\begin{equation}
    \label{eq-eom-th-1}
    \ud{}^*\ud\vartheta=\frac{\ma}{2\mb}\ud{}^*J,\text{ or, }\nabla_a\nabla^a\vartheta=-\frac{\ma}{4\mb}R_{abcd}{}^*R^{bacd}.
\end{equation}
Therefore, $S_3$ results in the same dynamics as the original one $S$, as long as one identifies $\nabla_a\vartheta$ with ${}^*\hat H/2\mb$.

\subsection{Asymptotic behaviors of the 2-form field}
\label{sec-afs}

Now, one can use the results in sec.~\ref{sec-cs} to compute the asymptotic behaviors of $\boldsymbol\theta^{\hat\mu}$, $\boldsymbol\omega_{\hat\mu}{}^{\hat\nu}$ and $\boldsymbol B$ without actually solving eqs.~\eqref{eq-eoms-fof}.
Since in this section, we are interested in the scalar memory, we will focus on the 2-form field $\boldsymbol B$.  
The asymptotic behaviors of $\boldsymbol\theta^{\hat\mu}$ and $\boldsymbol\omega_{\hat\mu}{}^{\hat\nu}$ are discussed in Appendix~\ref{sec-app-met}.

To fix the asymptotic behavior of $\boldsymbol B$, one notes that 
\begin{equation}
    \label{eq-idf}
    \ud\vartheta=\frac{{}^*\boldsymbol{\hat H}}{2\mb}=\frac{^*\ud\boldsymbol B+\ma\boldsymbol J}{2\mb},
\end{equation}
so taking the Hodge dual, one obtains\footnote{Note that there is a sign difference in defining the Hodge dual of a 1-form from refs.~\cite{Campiglia:2018see}.} 
\begin{equation}
    \label{eq-hexp}
    3\nabla_{[a}B_{bc]}=2\mb\epsilon^d{}_{abc}\nabla_d\vartheta-\ma{}^*J_{abc}.
\end{equation}
Then, usually, one wants to fix the gauge condition.
For example, ref.~\cite{Campiglia:2018see} chose Lorenz gauge condition $\nabla^bB_{ab}=0$. 
This cannot be made in the current work, because even if this condition is made at one moment, the equation of motion \eqref{eq-eom-b} for $\boldsymbol B$ may violate this condition at the later moment.
Note eq.~\eqref{eq-eom-b} can be rewritten as 
\begin{equation}
    \label{eq-eom-b-1}
    \nabla^c\nabla_cB_{ab}+2\nabla_{[a}\nabla^cB_{b]c}+R_{abcd}B^{cd}+2R_{[a}^dB_{b]d}+\ma\nabla^c{}^*J_{cab}=0.
\end{equation}
This is the equation of motion for $\boldsymbol B$, way more complicated than the one considered in ref.~\cite{Campiglia:2018see}.
On the other hand, one may consider the gauge condition proposed in ref.~\cite{Seraj:2021qja}. 
However, that gauge condition is not preserved under the BMS transformation.
So here, we will not impose any gauge condition.
One then formally expands the components of $B_{ab}$ \cite{Campiglia:2018see},
\begin{subequations}
    \begin{gather}
        B_{ur}=\frac{B_{ur}^{(1)}}{r}+\frac{B_{ur}^{(2)}}{r^2}+\order{\frac{1}{r^3}},\quad
        B_{uA}=B_{uA}^{(0)}+\order{\frac{1}{r}},\\
        B_{rA}=B_{rA}^{(0)}+\frac{B_{rA}^{(1)}}{r}+\order{\frac{1}{r^2}},\quad
        B_{AB}=r\hat\epsilon_{AB}\mathcal B+\order{r^0},
    \end{gather}
\end{subequations}
where all expansion coefficients are functions of $u$ and $x^A$.
By expanding eq.~\eqref{eq-hexp}, one can get
\begin{equation}
    \dot B_{rA}^{(0)}=0,\quad
    \vartheta_1=-\frac{1}{2\mb}\left(\mathcal B-\hat\epsilon^{AB}\mathscr D_AB_{rB}^{(0)}\right),\label{eq-th1-b}
\end{equation}
which can be compared with eq.~(4.5) in ref.~\cite{Campiglia:2018see}.
The equations for the remaining expansion coefficients are useless in the following discussion, so we do not write them\footnote{Interested Readers can refer to ref.~\cite{Campiglia:2018see}.}.
One could even consider more terms with positive powers of $r$, but their coefficients are irrelevant to $\vartheta_n$ (the series expansion coefficients of $\vartheta$).
Now, let $B_{rA}^{(0)}=\mathscr D_A\varsigma'+\hat\epsilon_{A}{}^B\mathscr D_B\varsigma$ with $\varsigma$ and $\varsigma'$ both $u$ independent.
Then, one has 
\begin{equation}
    \label{eq-th1-b-2}
    \vartheta_1=-\frac{1}{2\mb}(\mathcal B-\mathscr D^2\varsigma).
\end{equation}
This actually agrees with eq.~(3.12a) in ref.~\cite{Seraj:2021qja}.

\subsection{Large gauge transformations}
\label{sec-gtf}

As discussed in section~\ref{sec-cs}, Chern-Simons modified gravity in the first order formalism enjoys three types of symmetries: diffeomorphism, local Lorentz and gauge symmetries.
When one recedes from the source of gravity and approaches the infinity, one might expect these transformations tend to identity, as in many treatments. 
However, they can be nontrivial even at the infinity. 
In this subsection, nontrivial gauge transformations will be studied.
For asymptotic diffeomorphisms and local Lorentz transformations in the first-order formalism, please refer to Appendix~\ref{sec-app-bms}.

Let the gauge parameter $\mathfrak A_a$ have the following expansion,
\begin{subequations}
    \begin{gather}
        \mathfrak A_u=r\mathfrak A_u^{(-1)}+\mathfrak A_u^{(0)}+\order{r^{-1}},\\
        \mathfrak A_r=r\mathfrak A_r^{(-1)}+\mathfrak A_r^{(0)}+\frac{\mathfrak A_r^{(1)}}{r}+\order{\frac{1}{r^2}},\\
        \mathfrak A_A=r\mathfrak A_A^{(-1)}+\order{r^0},
    \end{gather}
\end{subequations}
Due to the boundary conditions of $B_{ab}$, one knows that $\mathfrak A^{(-1)}_r$ is constant, and 
\begin{equation}
    \dot{\mathfrak A}_r^{(0)}=\mathfrak A_u^{(-1)},\quad\dot{\mathfrak A}_A^{(-1)}=\mathscr D_A\mathfrak A_u^{(-1)}.
\end{equation}
Therefore, the expansion coefficients of $B_{ab}$ transform in the following way,
    \begin{equation}
        \delta_{\mathfrak A} B_{rA}^{(0)}=\mathfrak A_A^{(-1)}-\mathscr D_A\mathfrak A_r^{(0)},\quad 
        \delta_{\mathfrak A}\mathcal B=\hat\epsilon^{AB}\mathscr D_A\mathfrak A_B^{(-1)}.
    \end{equation}
One can check that $\delta_{\mathfrak A}\vartheta_1=0$ as expected.
The transformation laws for the remaining components of $B_{ab}$ are useless for the following discussion.
For the following discussion, decompose $\mathfrak A_A^{(-1)}=\mathscr D_A \varrho'+\hat\epsilon_{AB}\mathscr D^B\varrho$, then one has 
\begin{equation}
    \label{eq-gtf-2}
    \delta_\varrho\varsigma=\varrho,\quad\delta_\varrho\mathcal B=\mathscr D^2\varrho.
\end{equation}
The transformation of the electric part $(\varsigma')$ of $B_{rA}^{(0)}$ is irrelevant.
These can be used to calculate the conserved charge associated with the gauge transformations, and the discussion of the scalar memory effect in the next subsection.

\subsection{Covariant phase space structure and the scalar memory}
\label{sec-phas}

From the previous discussion, one could easily obtain the symplectic potential, 
\begin{equation}
    \label{eq-def-sym}
    \theta_3(\delta)=\int\left[\boldsymbol{\mathfrak L}_{\hat\mu}\wedge\delta\boldsymbol\theta^{\hat\mu}+\boldsymbol\Phi^{\hat\mu\hat\nu}\wedge\delta\boldsymbol\omega_{\hat\mu\hat\nu}+\frac{1}{4\mb}\,{}^*\boldsymbol{\hat H}\wedge\delta\boldsymbol B\right],
\end{equation}
with $\boldsymbol\Phi^{\hat\mu\hat\nu}=\kappa\boldsymbol\Sigma^{\hat\mu\hat\nu}+\frac{\ma}{4\mb}\,{}^*\boldsymbol{\hat H}\wedge\boldsymbol\omega^{\hat\mu\hat\nu}$.
Here, the subscript `3' indicates that this is for the action $S_3$.
Although the second term above is related to the tensor sector, we include it for completeness.
Therefore, the presymplectic is 
\begin{equation}
    \label{eq-def-psym}
        \Xi'_\Sigma(\delta_1,\delta_2)=\int_\Sigma\bigg[\delta_1\boldsymbol{\mathfrak L}_{\hat\mu}\wedge\delta_2\boldsymbol\theta^{\hat\mu}+\delta_1\boldsymbol\Phi^{\hat\mu\hat\nu}\wedge\delta_2\boldsymbol\omega_{\hat\mu\hat\nu}+\frac{1}{4\mb}\delta_1\,{}^*\boldsymbol{\hat H}\wedge\delta_2\boldsymbol B
        -\langle\delta_1\leftrightarrow\delta_2\rangle\bigg],
\end{equation}
where $\langle\delta_1\leftrightarrow\delta_2\rangle$ means to exchange $\delta_1 $ and $\delta_2$ in the previous terms.
One could evaluate the above expression on the null infinity $\scri$, and it turns out that 
\begin{equation}
    \label{eq-ev-psym-1}
    \begin{split}
    \Xi'_{\scri}(\delta_1,\delta_2)=&\int\ud u\ud^2x\sqrt{\gamma} \left( \frac{\kappa}{2}\delta_1c^{A}_{B}\delta_2N_{A}^{B}-\frac{1}{2}\delta_1\mathcal B\delta_2N-\langle\delta_1\leftrightarrow\delta_2\rangle\right),\\
        =&\int\ud u\ud^2x\sqrt{\gamma} \left( \frac{\kappa}{2}\delta_1c^{A}_{B}\delta_2N_{A}^{B}+\mb\delta_1\vartheta_1\delta_2N-\frac{1}{2}\mathscr D^2\delta_1\varsigma \delta_2N-\langle\delta_1\leftrightarrow\delta_2\rangle\right),
    \end{split}
\end{equation}
Note that $\varsigma$ is independent of $u$. 
If one calculates the presymplectic current over a patch $\mathscr B$ of $\scri$ as in eq.~\eqref{eq-flux-v}, one has 
\begin{equation}
    \label{eq-ev-psym-b}
    \begin{split}
    \Xi'_{\mathscr B}(\delta_1,\delta_2)=&\int\ud u\ud^2x\sqrt{\gamma} \left( \frac{\kappa}{2}\delta_1c^{A}_{B}\delta_2N_{A}^{B}+\mb\delta_1\vartheta_1\delta_2N\right)
    -\frac{1}{2}\int\ud^2x\sqrt\gamma\mathscr D^2\delta_1\varsigma\delta_2\Delta\vartheta_1\\
    & -\langle\delta_1\leftrightarrow\delta_2\rangle.
    \end{split}
\end{equation}
This expression is similar to eq.~(3.19) in ref.~\cite{Seraj:2021qja}.
From these expressions, $\mathcal B$ is conjugate to $N$. 
Indeed, $\dot{\mathcal B}=-2\mb N$ according to eq.~\eqref{eq-th1-b}.

To compute the conserved charges for the gauge symmetry, one considers  the gauge transformation $\delta_{\mathfrak A}\boldsymbol B=\ud\boldsymbol{\mathfrak A}$, that is, eq.~\eqref{eq-gtf-2}.
Of course, the remaining quantities in eq.~\eqref{eq-ev-psym-1} stay the same.
So 
\begin{equation}
    \label{eq-sypre-gtf}
    \delta F_{\mathfrak A}\equiv\Xi'_{\mathscr B}(\delta,\delta_{\mathfrak A})=\frac{1}{2}\int\ud u\ud^2x\sqrt\gamma(\delta N)\mathscr D^2\varrho,
\end{equation}
on-shell.
This is integrable,  and the Hamiltonian, i.e., the flux is, 
\begin{equation}
    \label{eq-flux-g}
    F_{\varrho}=\frac{1}{2}\int \ud u\ud^2x\sqrt{\gamma}N\mathscr D^2\varrho,
\end{equation}
and the Noether charge is
\begin{equation}
    \label{eq-hal}
    Q_{\varrho}=-\frac{1}{2}\int\ud^2x\sqrt{\gamma}\vartheta_1\mathscr D^2\varrho,
\end{equation}
where the subscripts are changed to $\varrho$ for simplicity.
Apparently, the flux-balance law is satisfied,
\begin{equation}
    \label{eq-fbl}
    F_{\varrho}[\mathscr B]=-(Q_{\varrho}[\mathscr C_2]-Q_{\varrho}[\mathscr C_1]).
\end{equation}
Although not explicit at the level of the action $S_3$, the presymplectic current \eqref{eq-ev-psym-1} gives a new conserved charge under the shift $\vartheta_1\rightarrow\vartheta_1+\iota$ with $\iota$ a constant,
\begin{equation}
    \label{eq-cc}
    Q_\iota=\int\ud^2x\sqrt{\gamma}\mb\vartheta_1,
\end{equation}
whose corresponding flux is $F_\iota[\mathscr B]=-(Q_\iota[\mathscr C_2]-Q_\iota[\mathscr C_1])=-\int\ud u\ud^2x\sqrt{\gamma}\mb N$.
This is the scalar charge usually talked about.

Although the scalar memory cannot be detected with laster interferometers, pulsar timing arrays or the Gaia mission, it can still be defined to be the lasting change in the radiative degree of freedom in the scalar sector before and after the passage of the scalar gravitational wave.
The states before and after the passage of the scalar wave are both vacua with $\mathcal B$ independent of $u$.
So in the theory defined by the presymplectic current \eqref{eq-ev-psym-1}, the vacua is characterized by $\mathcal B$. 
The change in $\mathcal B$ measures the scalar memory.
One can compute the following Poisson brackets,
\begin{equation}
    \label{eq-pobr}
    \{Q_\varrho,\mathcal B\}=\mathscr D^2\varrho,\quad\{Q_\iota,\mathcal B\}=-2\mb,
\end{equation}
due to the basic Poisson bracket $\{\mathcal B(u,x),N(u',x')\}=-\frac{2}{\sqrt\gamma}\delta(u-u')\delta^2(x-x')$.
So the change in $\mathcal B$ can be generated by the linear combination of the two kinds of infinitesimal transformation $\iota Q_\iota+Q_\varrho$, i.e., \cite{Seraj:2021qja}
\begin{equation}
    \label{eq-cb-sm}
    \Delta\mathcal B=\{\iota Q_\iota+Q_\varrho,\mathcal B_0\}.
\end{equation}
with $\mathcal B_0$ before the arrival of the scalar wave.
This means that the scalar memory can also be interpreted  as the vacuum transition.
$\Delta \mathcal B$ is constrained by the flux-balance laws $\iota F_\iota=-\iota\Delta Q_\iota$, $F_\varrho=-\Delta Q_\varrho$, that is,
\begin{equation}
    \label{eq-cst-b}
    \int\ud^2x\sqrt\gamma\varrho\mathscr D^2\Delta\mathcal B=-4\mb\Delta Q_\varrho,\quad \int\ud^2x\sqrt\gamma\iota\Delta\mathcal B|_{l=0}=-2\iota\Delta Q_\iota.
\end{equation}
Formally, they are similar to those constraints in section~\ref{sec-cst-mm}.

\section{Conclusion}
\label{sec-con}

In this work, we analyzed the asymptotically flat spacetime in Chern-Simons modified gravity with Penrose's conformal completion method. 
This justifies the boundary conditions imposed in ref.~\cite{Hou:2021cbd}.
It also helps identify the asymptotic structures on the null infinity and the BMS transformations.
We find out that the asymptotic structures and the BMS transformations are very similar to those in Brans-Dicke theory \cite{Hou:2020wbo}. 
This is because the interaction term [the second one in eq.~\eqref{eq-cs-act}] is of higher order in $1/r$, and thus can be ignored in analyzing the asymptotic structure. 
It also implies the absence of the parity violation effect in memories.
Then, the conserved charges and fluxes are computed for the BMS algebra with Wald-Zoupas formalism.
The flux-balance laws actually take the similar forms as those in Brans-Dicke theory, for the same reason explained above.
These laws are rewritten to constrain displacement, spin and CM memories in the tensor sector, as in general relativity and Brans-Dicke theory, e.g., see refs.~\cite{Compere:2019gft,Flanagan:2015pxa,Nichols:2017rqr,Nichols:2018qac,Hou:2020wbo,Tahura:2020vsa}, et al.
These constraints can be used to predict the magnitudes of memory effects, and evaluate their possibilities detection by interferometers, pulsar timing arrays and Gaia mission.

The memory effect in the scalar sector is then discussed. 
For this purpose, the original theory is rewritten in the first-order formalism, and the Chern-Simons scalar field $\vartheta$ is dual to a differential 2-form field $B_{ab}$, following refs.~\cite{Campiglia:2017dpg,Campiglia:2018see,Seraj:2021qja}.
The new theory possesses the gauge symmetry associated with  $B_{ab}$, and this symmetry is nontrivial at the null infinity. 
The vacuum states in the scalar sector are thus also degenerate, and they transform to each other via the large gauge transformation.
This causes the scalar memory, just like the displacement memory \cite{Strominger:2014pwa,Strominger:2018inf}. 
This memory is also constrained by the flux-balance laws associated with the large gauge transformation.
Although Chern-Simons theory is way more complicated than Brans-Dicke theory, the analysis of the scalar memory using the dual formalism is still similar to that in ref.~\cite{Seraj:2021qja}. 
This is also due to the fact that in the infrared regime, all interaction becomes weak enough.

In the future, the magnitudes of memory effects will be computed using the constraints obtained in this work, just like refs.~\cite{Nichols:2017rqr,Nichols:2018qac,Tahura:2021hbk}. 
One could also extend this analysis to the remaining modified theories of gravity, such as massive gravities \cite{deRham:2014zqa}, Einstein-\ae ther theory \cite{Jacobson:2000xp,Jacobson:2004ts}, Ho\v rava-Lifshitz gravity \cite{Horava:2009uw}, and so on.
The local Lorentz invariance is broken in Einstein-\ae{}ther theory and Ho\v rava-Lifshitz gravity, so gravitational waves might travel at superluminal speeds \cite{Gong:2018cgj,Gong:2018vbo}.
Radiative modes of these theories would arrive at the future timelike infinity or spacelike infinity, instead of null infinity.
It is interesting to study whether these theories also predict memory effects, and if so, their properties.

\acknowledgments
This work was supported by the National Natural Science Foundation of China under grant Nos.~11633001 and 11920101003, and the Strategic Priority Research Program of the Chinese Academy of Sciences, grant No.~XDB23000000.
Tao Zhu is supported in part by the National Key Research and Development Program of China Grant No.~2020YFC2201503, the Zhejiang Provincial Natural Science Foundation of China under Grants No.~LR21A050001 and No.~LY20A050002.

\appendix

\section{Tetrads and connection in the first-order formalism}
\label{sec-app-met}

In this appendix, the asymptotic behaviors of the tetrads and connection will be presented.
For this purpose, one should first determine a suitable tetrad basis.
One notes that in eq.~\eqref{eq-bc}, $e^{2\beta}r^{-1}V\ud u^2-2e^{2\beta}\ud u\ud r=-2e^{2\beta}\ud u[\ud r-(2r)^{-1}V\ud u]$, so let us define
\begin{equation}
    \label{eq-def-ln}
    l_a=-e^\beta(\ud u)_a,\quad n_a=e^{\beta}\left[\frac{V}{2r}(\ud u)_a-(\ud r)_a\right].
\end{equation}
Next, following ref.~\cite{Madler:2016xju}, introduce a complex polarization dyad $m^a$ with $m^a=(0,0,m^A)$ such that $h^{AB}=m^A\bar m^B+\bar m^Am^B$, where bar implies to take the complex conjugate.
The corresponding covector is 
\begin{equation}
    \label{eq-def-m}
    m_a=g_{ab}m^b=h_{AB}m^A[(\ud x^B)_a-U^B(\ud u)_a].
\end{equation}
These covectors $l_a,\,n_a,\,m_a,$ and $\bar m_a$ form the Newman-Penrose basis \cite{Newman:1961qr}, and we simply identify $\boldsymbol\theta^{\hat\mu}$ with them in the following way,
\begin{equation}
    \label{eq-def-np}
    \boldsymbol\theta^{\hat 0}=\boldsymbol l,\quad \boldsymbol\theta^{\hat 1}=\boldsymbol n,\quad \boldsymbol\theta^{\hat 2}=\boldsymbol m,\quad \boldsymbol\theta^{\hat 3}=\bar{\boldsymbol m}.
\end{equation}
So the complex conjugation amounts to exchanging $\hat 2$ and $\hat 3$.
The vector basis includes 
\begin{equation}
    \label{eq-def-np-v}
    (e_{\hat0})^a=-n^a,\quad (e_{\hat1})^a=-l^a,\quad \bvec{e}{2}{a}=\bar m^a,\quad \bvec{e}{3}{a}=m^a.
\end{equation}
In this basis, the metric takes the following matrix form,
\begin{equation}
    \label{eq-met-mat}
    g_{\hat\mu\hat\nu}=\left(
        \begin{array}{cccc}
            0 & -1 &0  &0 \\
            -1 & 0 &0 & 0\\
            0 & 0 & 0 & +1\\
            0 & 0 & +1 & 0
        \end{array}
    \right).
\end{equation}
To obtain the asymptotic behavior of $m_a$, one first assumes 
\begin{equation}
    \label{eq-mas}
    m^A=\frac{\gamma^A}{r}+\frac{\mc^A}{r^2}+\frac{\md^A}{r^3}+\order{\frac{1}{r^4}},
\end{equation}
with $\gamma^A$, $\mc^A$ and $\md^A$  arbitrary vectors independent of $r$, and $\gamma^{AB}=\gamma^A\bar\gamma^B+\bar\gamma^A\gamma^B$.
One can also finds out that    $\hat\epsilon_{AB}=-i\gamma_{A}\wedge\bar\gamma_{B}$.
$\gamma^A$ and $\bar\gamma^A$ are independent of $u$, and serve as basic vectors. 
So let $\mc^A=-(\mc\bar\gamma^A+\mc'\gamma^A)/2$, and $\md^A=-(\md\bar\gamma^A+\md'\gamma^A)/2$, i.e., $\mc,\,\mc',\,\md$, and $\md'$  are components. 
Since $h^{AB}=\gamma^{AB}/r^2-c^{AB}/r^3+\gamma^{AB}c_C^Dc_D^C/4r^4+\order{r^{-5}}$, one knows that
\begin{subequations}
\begin{gather}
    c^{AB}=\bar\mc\gamma^A\gamma^B+\mc\bar\gamma^A\bar\gamma^B,\quad \mc'+\bar\mc'=0,\label{eq-cinc}\\
     \md=\frac{\bar\mc'\mc}{2},\quad \md'+\bar\md'=\frac{\mc'\bar\mc'-\mc\bar\mc}{2}.
\end{gather}
\end{subequations}
In addition, the last expression in eq.~\eqref{eq-exp-m} is usually required, which is equivalent to
\begin{equation}
    \label{eq-det-con}
    m^A\wedge\bar m^B=\frac{1}{r^2}\gamma^A\wedge\bar\gamma^B,
\end{equation}
which does impose any new constraints.

In fact, there is freedom in choosing $\gamma^A$ and $\bar\gamma^A$.
A spin, $\gamma^A\rightarrow e^{i\chi}\gamma^A$ and $\bar\gamma^A\rightarrow e^{-i\chi}\bar\gamma^A$, gives a new basis.
With this freedom, it is possible to choose a spin $m^A\rightarrow e^{i \chi_1/r}m^A$ with $\chi_1=-i\mc'/2$ to remove $\mc'$, and so $\md=0$.
Then, another spin $m^A\rightarrow e^{i\chi_2/r^2}m^A$ with $\chi_2=i(\bar\md'-\md')/4$ annihilates the imaginary part of $\md'$, so that $\md'=-\mc\bar\mc/4$. 
In the following, we will work in this gauge.
Let $W_{B_1\cdots B_pC_{1}\cdots C_{q}}$ be an arbitrary ($0,p+q$)-type tensor defined on the unit 2-sphere, then the component $W=\gamma^{B_1}\cdots\gamma^{B_p}\bar\gamma^{C_{1}}\cdots\bar\gamma^{C_{q}}W_{B_1\cdots B_pC_{1}\cdots C_{q}}$ transforms to $e^{is\psi}W$ under the spin with $s=p-q$.
The number $s$ is called the spin weight of the component $W$.
One can check that $\mc$  has spin weight of 2.
Now, introduce derivatives $\eth$ and $\bar\eth$, which act on $W$ in the following way,
\begin{gather}
    \eth W=\gamma^A\gamma^{B_1}\cdots\gamma^{B_p}\bar\gamma^{C_1}\cdots\bar\gamma^{C_q}\mathscr D_AW_{B_1\cdots B_pC_1\cdots C_q}=\gamma^A\pd_AW-s\zeta W,\\
    \bar\eth W=\bar\gamma^A\gamma^{B_1}\cdots\gamma^{B_p}\bar\gamma^{C_1}\cdots\bar\gamma^{C_q}\mathscr D_AW_{B_1\cdots B_pC_1\cdots C_q}=\bar\gamma^A\pd_AW+s\bar\zeta W,
\end{gather}
respectively, where $\zeta=\mathscr D_A\gamma^A$.
It is easy to realize that they are equivalent to the $\eth$ and $\bar\eth$ defined in refs.~\cite{Goldberg:1966uu,Penrose:1985bww,JStewart1991}, although here, they are defined with respect to an arbitrary basis $\gamma^A$ and $\bar\gamma^A$. 

So, with the solutions~\eqref{eq-met-sols}, one easily finds out that
\begin{gather*}
    l_a=-\left[ 1+\frac{\beta_2}{r^2} +\order{\frac{1}{r^3}}\right](\ud u)_a,\\
    n_a=\left[ -\frac{1}{2}+\frac{m}{r} +\order{\frac{1}{r^2}}\right](\ud u)_a- \left[ 1+\frac{\beta_2}{r^2} +\order{\frac{1}{r^3}}\right](\ud r)_a,\\
    m_a=\left[ \frac{\bar\eth\mc}{2r}+ \frac{1}{r^2} \left( \frac{2}{3}\mn -\frac{\bar\mc\eth\mc}{8}-\frac{9\mc\eth\bar\mc}{24} \right) +\order{\frac{1}{r^3}}\right](\ud u)_a
    +\left[  r\gamma_A+\frac{\mc}{2}\bar\gamma_A +\order{\frac{1}{r}}\right](\ud x^A)_a,
\end{gather*}
where $\mn=\gamma^AN_A$.
Taking the complex conjugate of $m_a$, one obtains the asymptotic behavior of $\bar m_a$.
Due to the symmetry of $\boldsymbol\omega_{\hat\mu}{}^{\hat\nu}$, one only has to determine the asymptotic behaviors of $\boldsymbol\omega_{\hat 0}{}^{\hat0},\,\boldsymbol\omega_{\hat0}{}^{\hat2},\,\boldsymbol\omega_{\hat1}{}^{\hat2},$ and $\boldsymbol\omega_{\hat2}{}^{\hat2}$.
The remaining components are related to these or zero: $\boldsymbol\omega_{\hat1}{}^{\hat1}=-\boldsymbol\omega_{\hat0}{}^{\hat0}$, $\boldsymbol\omega_{\hat3}{}^{\hat1}=\bar{\boldsymbol\omega}_{\hat0}{}^{\hat3}=\bar{\boldsymbol\omega}_{\hat2}{}^{\hat1}=\boldsymbol\omega_{\hat0}{}^{\hat2}$, $\bar{\boldsymbol\omega}_{\hat1}{}^{\hat3}=\boldsymbol\omega_{\hat3}{}^{\hat0}=\bar{\boldsymbol\omega}_{\hat2}{}^{\hat0}=\boldsymbol\omega_{\hat1}{}^{\hat2}$, $\boldsymbol\omega_{\hat3}{}^{\hat3}=-\boldsymbol\omega_{\hat2}{}^{\hat2}$ and $\boldsymbol\omega_{\hat0}{}^{\hat1}=\boldsymbol\omega_{\hat1}{}^{\hat0}=\boldsymbol\omega_{\hat2}{}^{\hat3}=\boldsymbol\omega_{\hat3}{}^{\hat2}=0$.
One can check that $\boldsymbol\omega_{\hat 0}{}^{\hat0}$ is real, $\boldsymbol\omega_{\hat2}{}^{\hat2}$ imaginary, and $\boldsymbol\omega_{\hat0}{}^{\hat2},\,\boldsymbol\omega_{\hat1}{}^{\hat2}$ complex, so there are 6 real degrees of freedom.
The basic four connection 1-forms are\footnote{Note that $\omega_{\hat 1}{}^{\hat2}{}_a(\pd_r)^a=0$ in this approximation.}
\begin{subequations}
    \label{eq-ab-ome}
    \begin{gather}
        \begin{split}
            \omega_{\hat0}{}^{\hat0}{}_a=&\left\{\frac{1}{r^2}\left[ m+\frac{\mb}{8\kappa}\vartheta_1N+\frac{\pd_u(\mc\bar\mc)}{16} \right]+\order{\frac{1}{r^3}}\right\}(\ud u)_a+\left[\frac{1}{8r^3}\left(\frac{\mb}{\kappa}\vartheta_1^2+\mc\bar\mc\right)\right.\\
            &\left.+\order{\frac{1}{r^4}}\right](\ud r)_a
            +\left[\frac{1}{2r}(\gamma_A\eth\bar\mc+\bar\gamma_A\bar\eth\mc)+\order{\frac{1}{r^2}}\right](\ud x^A)_a,
        \end{split}\\
        \begin{split}
            \omega_{\hat0}{}^{\hat2}{}_a=&\left[\frac{1}{r^2} \left( \frac{\dot\mc\eth\bar\mc}{4}-\eth m \right)+\order{\frac{1}{r^3}}\right] (\ud u)_a+\left[\frac{\bar\eth\mc}{2r^2}+\order{\frac{1}{r^3}}\right](\ud r)_a\\
            &+\left[\frac{1}{2}(\bar\gamma_A\dot\mc-\gamma_A)+\order{\frac{1}{r}}\right](\ud x^A)_a,
        \end{split}\\
        \begin{split}
            \omega_{\hat 1}{}^{\hat2}{}_a=\left[-\frac{1}{r^3} \left( \frac{\mn}{3}+\frac{\mb}{8\kappa}\vartheta_1\eth\vartheta_1 \right)+\order{\frac{1}{r^4}}\right](\ud u)_a
            +\left[ 1+\order{\frac{1}{r^2}} \right]\gamma_A(\ud x^A)_a,
        \end{split}\\
        \begin{split}
        \omega_{\hat2}{}^{\hat2}{}_a=&\left\{ \frac{1}{4r^2}\left[ \eth\eth\bar\mc-\bar\eth\bar\eth\mc +\frac{1}{2}(\mc\dot{\bar\mc}-\bar\mc\dot\mc)\right] +\order{\frac{1}{r^3}}\right\} (\ud u)_a+ \left[ \order{\frac{1}{r^5}} \right](\ud r)_a\\
            &+\left[(\bar\gamma_A\zeta-\gamma_A\bar\zeta)+\frac{1}{2r}(\gamma_A\eth\bar\mc-\bar\gamma_A\bar\eth\mc)+\order{\frac{1}{r^2}}\right](\ud x^A)_a,
        \end{split}
    \end{gather}
\end{subequations}
where only the leading terms in $1/r$ are included.

\section{BMS transformations in first-order formalism}
\label{sec-app-bms}

Let us briefly discuss the asymptotic symmetries that preserve the boundary conditions of the tetrad basis.
Suppose the vector field $\xi^a$ generates an infinitesimal asymptotic symmetry.
Its action on the tetrad basis is given by 
\begin{equation}
    \label{eq-tet-bms}
    \delta_\xi\boldsymbol{\theta}^{\hat\mu}=\lie_\xi\boldsymbol{\theta}^{\hat\mu}+\lambda^{\hat\mu}{}_{\hat\nu}\boldsymbol{\theta}^{\nu},
\end{equation}
where $\lambda^{\hat\mu}{}_{\hat\nu}$ represents the infinitesimal transformation preserving the metric \eqref{eq-met-mat}.
Although we are using the Newman-Penrose tetrads instead of the orthonormal tetrads, we still call the transformation generated by $\lambda^{\hat\mu}{}_{\hat\nu}$ the infinitesimal local Lorentz transformation.
This transformation is induced by $\xi^a$.
The inclusion of the second term in eq.~\eqref{eq-tet-bms} is necessary, otherwise, one would get a very constrained transformation.
One can also check that $\lambda^{\hat\mu}{}_{\hat\nu}$ satisfies $\lambda_{\hat\mu\hat\nu}=g_{\hat\mu\hat\rho}\lambda^{\hat\rho}{}_{\hat\nu}=-\lambda_{\hat\nu\hat\mu}$.
Now, let us determine $\xi^a$.
Since $\xi^a$ preserves the boundary conditions of $\boldsymbol\theta^{\hat\mu}$, then $\delta_\xi l_a$ does not contain $r$ component, 
\begin{equation}
    \label{eq-xi-u}
    \delta_\xi l_r=\lie_\xi l_r+\lambda^{\hat0}{}_{\hat1}n_r=\lie_\xi l_r=l_u\frac{\pd \xi^u}{\pd r}=0,
\end{equation}
where $\lambda^{\hat0}{}_{\hat1}=-\lambda_{\hat1\hat1}=0$.
Therefore, there exists a function $f=f(u,x^A)$ such that $\xi^u=f$.
Similarly, $\delta_\xi n_A=\delta_\xi l_A=0=\delta_\xi m_r$, that is,
\begin{subequations}
    \begin{gather}
        e^\beta\left(\frac{V}{2r}\pd_Af-\pd_A\xi^r\right)+\bar\lambda^{\hat2}{}_{\hat0}m_A+\lambda^{\hat2}{}_{\hat0}\bar m_A=0,\\
        -e^\beta\pd_Af+\bar\lambda^{\hat2}{}_{\hat1}m_A+\lambda^{\hat2}{}_{\hat1}\bar m_A=0,\\
        m_A\pd_r\xi^A-e^\beta\lambda^{\hat2}{}_{\hat1}=0,\label{eq-dmr}
    \end{gather}
\end{subequations}
where $m_A=h_{AB}m^B$.
Combining these relations together leads to 
\begin{gather}
    \lambda^{\hat2}{}_{\hat0}=e^\beta m^A\left(\pd_A\xi^r-\frac{V}{2r}\pd_Af\right),\\
    \lambda^{\hat2}{}_{\hat1}=e^\beta m^A\pd_Af,\\
    \pd_r\xi^A=-e^{2\beta}h^{AB}\pd_Bf,\quad\text{ i.e., }\quad\xi^A=Y^A-\pd_Bf\int_r^\infty e^{2\beta}h^{AB},\label{eq-xi-A}
\end{gather}
for some vector field $Y^A=Y^A(u,x^B)$.
One should also make sure eq.~\eqref{eq-det-con} remains unchanged, so 
\begin{equation}
    \label{eq-xi-r}
    \xi^r=\frac{r}{2}(U^A\pd_Af-\mathscr D_A\xi^A).
\end{equation}
With the asymptotic solutions in the previous subsection, one finds out that
\begin{subequations}
    \begin{gather}
        \xi^A=Y^A-\frac{\mathscr D^Af}{r}+\frac{c^{AB}\mathscr D_Bf}{2r^2}
        -\frac{1}{r^3}\left( \frac{c_{B}^Cc^{B}_C}{16}-\frac{\mb}{24\kappa}\vartheta_1^2 \right)\mathscr D^Af+\order{\frac{1}{r^4}},\\
        \xi^r=-\frac{r}{2}\psi+\frac{1}{2}\mathscr D^2f-\frac{1}{2r}\bigg[ (\mathscr D_Af)\mathscr D_Bc^{AB}
        +\frac{1}{2}c^{AB}\mathscr D_A\mathscr D_Bf \bigg]+\order{\frac{1}{r^2}},
    \end{gather}
where $\psi=\mathscr D_AY^A$.
Similarly, one also knows that 
\begin{gather}
    \lambda^{\hat2}{}_{\hat0}=-\frac{\eth\psi}{2}+\frac{1}{2r} \left[ \eth(\mathscr D^2+1)f+\frac{1}{2}\mc\bar\eth\psi \right]+\order{\frac{1}{r^2}},\\
    \lambda^{\hat2}{}_{\hat1}=\frac{\eth f}{r}-\frac{\mc\bar\eth f}{2r^2}+\order{\frac{1}{r^3}}.
\end{gather}
\end{subequations}
Up to this point, one has not yet determined the properties of $f$ and $Y^A$, which requires to make use of the remaining variations of the tetrad components.
With the sum of  $\delta_\xi l_u=\order{r^{-2}},\,\delta_\xi n_r=\order{r^{-2}}$, one knows that $\dot f=\psi/2$, while the difference between $\delta_\xi l_u$ and $\delta_\xi n_r$ results in,
\begin{equation}
    \label{eq-lam00}
    \lambda^{\hat0}{}_{\hat0}=-\frac{\psi}{2}+\order{r^{-2}}.
\end{equation}
Also, consider $\delta_\xi m_u+U^A\delta_\xi m_A=\order{1/r}$ to lead to 
\begin{equation}
    \label{eq-ytd}
    \dot Y^A=0,
\end{equation}
that is, $Y^A$ is a function of the angular coordinates, and so 
\begin{equation}
    \label{eq-f-a}
    f=\alpha+\frac{u}{2}\psi,
\end{equation}
where $\alpha=\alpha(x^A)$ is arbitrary.
Finally, $\delta_\xi m_A=\order{r}$ implies that 
\begin{gather}
    \lie_Y\gamma_{AB}=\psi\gamma_{AB},\label{eq-cfkc}\\
    \lambda^{\hat2}{}_{\hat2}=\frac{1}{2}(\eth\bar\my-\bar\eth\my)+\order{\frac{1}{r}},
\end{gather}
where $\my=\gamma_AY^A$.
Indeed, $Y^A$ is a conformal Killing vector field on the unit 2-sphere.
In terms of $\my$, $\psi=\bar\eth\my+\eth\bar\my$, and eq.~\eqref{eq-cfkc} is equivalent to $\eth\my=0$.
One can also check that this $\xi^a$ also preserves the asymptotic behaviors of $\boldsymbol\omega_{\hat\mu}{}^{\hat\nu}$ and $\boldsymbol B$.

In summary, the generator of the asymptotic symmetry, determined by requiring the asymptotic behaviors of tetrad basis be unchanged, takes the exactly the same form as the one in the metric formalism \cite{Hou:2021cbd}.
Therefore, the asymptotic symmetry is still the Bondi-Metzner-Sachs symmmetry, or the extended one \cite{Flanagan:2015pxa}.
The results presented here are very similar to those in ref.~\cite{Godazgar:2018vmm}.

\bibliographystyle{JHEP}
\bibliography{memorycs_torsionful_dual_v2.bbl}

\end{document}